\documentclass{ws-ijmpa}
\usepackage[super,compress]{cite}
\usepackage{graphicx}
\usepackage{graphicx}
\usepackage{dcolumn}
\usepackage{bm}
\usepackage{epstopdf}
\newcommand{\hs}{\hspace*{0.5cm}}

\newcommand{\be}{\begin{equation}}
\newcommand{\ee}{\end{equation}}
\newcommand{\bea}{\begin{eqnarray}}
\newcommand{\eea}{\end{eqnarray}}
\newcommand{\ben}{\begin{enumerate}}
\newcommand{\een}{\end{enumerate}}
\newcommand{\bde}{\begin{widetext}}
\newcommand{\ede}{\end{widetext}}
\newcommand{\nn}{\nonumber}
\newcommand{\crn}{\nonumber \\}
\newcommand{\Tr}{\mathrm{Tr}}

\newcommand{\al}{\alpha}
\newcommand{\la}{\lambda}

\newcommand{\ga}{\gamma}

\newcommand{\om}{\omega}

\newcommand{\+}{\dagger}
\newcommand{\fr}{\frac}
\newcommand{\bc}{\begin{center}}
\newcommand{\ec}{\end{center}}
\newcommand{\Ga}{\Gamma}
\newcommand{\de}{\delta}

\newcommand{\ep}{\epsilon}

\newcommand{\La}{\Lambda}
\newcommand{\si}{\sigma}

\begin{document}
\catchline{}{}{}{}{}

\title{THE $D_4$ FLAVOR SYMMETRY IN 3-3-1 MODEL\\ WITH NEUTRAL LEPTONS}

\author{V. V. VIEN }

\address{Department of Physics, Tay Nguyen University, 567 Le
Duan, Buon Ma Thuot, Vietnam\\
wvienk16@gmail.com}

\author{H. N. LONG}

\address{Institute of Physics,
VAST, 10 Dao Tan, Ba Dinh, Hanoi, Vietnam\\
hnlong@iop.vast.ac.vn}

\maketitle

\begin{history}
\received{Day Month Year}
\revised{Day Month Year}
\end{history}

\begin{abstract}
We construct a $D_4$ flavor model based on $\mathrm{SU}(3)_C
\otimes \mathrm{SU}(3)_L \otimes \mathrm{U}(1)_X$ gauge symmetry
responsible for fermion masses and mixings. The neutrinos get
small masses from antisextets which are in a singlet and a doublet
under $D_4$. If the $D_4$ symmetry is violated as perturbation by
a Higgs triplet under $SU(3)_L$ and lying in $\underline{1}'''$ of
$D_4$, the corresponding neutrino mass mixing matrix gets the most
general form. In this case, the model can fit the experimental
data in 2012 on neutrino masses and mixing. Our results show that
the neutrino masses are naturally small and a little deviation from the
tribimaximal neutrino mixing form can be realized. The quark
masses and mixing matrix are also discussed. In the model
under consideration, the CKM matrix can be different from the unit
matrix. The scalar potential of the model is more simpler than
those of the model based on $S_3$ and $S_4$. Assignation of
VEVs to antisextets leads to the mixing of the new gauge bosons and those in the
Standard Model. The mixing in the charged gauge bosons as well as the neutral
gauge boson is considered.

\end{abstract}

\ccode{PACS numbers: 14.60.Pq, 14.60.St, 12.60.Fr, 11.30.Er}

\section{\label{intro}Introduction}

Following the discovery of neutrino oscillations, there has been a
considerable progress in determining values for neutrino mass
square differences $m^2_i- m^2_j$ and the mixing angles
relating mass eigenstates to flavor eigenstates. The most recent
fits suggest that one of the mixing angles is approximately zero
and another has a value that implies a mass eigenstate
that is nearly an equal mixture of $\nu_\mu$ and $\nu_\tau$. The
data in PDG2012 \cite{PDG2012, PDG1, PDG2, PDG3, PDG4}   imply: \bea
&&\sin^2(2\theta_{12})=0.857\pm 0.024 \hs
(t_{12}\simeq0.6717),\crn
&&\sin^2(2\theta_{13})=0.098\pm 0.013
\hs (s_{13}\simeq 0.1585),\crn && \sin^2(2\theta_{23})> 0.95,
\crn && \Delta m^2_{21}=(7.50\pm0.20)\times 10^{-5}
\mathrm{eV}^2,\hs \Delta m^2_{32}=(2.32^{+0.12}_{-0.08})\times
10^{-3}\mathrm{eV}^2.\label{PDG2012}\eea
These large neutrino mixing angles are completely different from the quark
 mixing ones defined by the Cabibbo- Kobayashi-Maskawa (CKM) matrix \cite{CKM, CKM1}   .
 This has stimulated work on flavor symmetries and non-Abelian discrete
 symmetries are considered to be the most attractive candidate to
formulate dynamical principles that can lead to the flavor
mixing patterns for quarks and lepton. There are many recent models based on
the non-Abelian discrete symmetries, such as $A_4$ ~\cite{A41, A42, A43, A44, A45,
A46, A47, A48, A49, A410, A411, A412, A413, A414, A415, A416, A417, A418}  ,
$A_5$ \cite{A51, A52, A53, A54, A55, A56, A57, A58, A59, A510, A511, A512, A513} ,
$S_3$\cite{S31,S32,S33,S34,S35,S36,S37,S38,S39,S310,S311,S312,S313,
S314,S315,S316,S317,S318,S319,S320,S321,S322,S323,S324,S325,
S326,S327,S328,S329,S330,S331,S332,S333,S334,S335,S336,
S337,S338,S339,S340,S341,S342} , $S_4$ \cite{S41,S42,S43,S44,
S45,S46,S47,S48,S49,S410,S411,S412,S413,
S414,S415,S416,S417,S418,S419,S420,S421,S422,S423,S424,S425,
S426,S427,S428,S429} , $D_4$ \cite{D41,D42,D43,D44,D45,D46,D47,
D48,D49,D410,D411,D412}   , $D_5$ \cite{D51, D52}   ,
$T'$ \cite{Tp1,Tp2,Tp3,Tp4,Tp5} and so forth. An alternative  extension of  the Standard Model (SM)
 is the 3-3-1 models, in which
  the SM gauge group $SU(2)_L\otimes U(1)_Y$ is extended to
  $SU(3)_L\otimes U(1)_X$,
 has been  investigated in Refs. \citen{331m1,331m2,331m3,331m4,331m5, 331r1,
 331r2, 331r4, 331r5, 331r6, e3311, e3312, e3313, e3314}. The anomaly
cancelation and the QCD asymptotic freedom  in the models require
that the number of fermion families is 3, and one family of quarks
has to transform under $\mathrm{SU}(3)_L$ differently from the two
others. In our previous works \cite{dlshA4, dlsvS4, dlnvS3}   , the discrete symmetries
have been explored to the 3-3-1 models. The simplest explanation is probably due to a
$S_3$ flavor symmetry which is the smallest non-Abelian discrete
group,  has been explored in our previous work \cite{dlnvS3} .
 In Ref. \citen{dlsvS4} we have studied the 3-3-1 model with neutral fermions based
    on $S_4$ group, in which most of the Higgs multiplets are in triplets under $S_4$
    except $\chi$ lying  in a singlet, and the exact tribimaximal
    form \cite{hps1,hps2,hps3,hps4} is obtained,
     where $\theta_{13}= 0$.

As we know, the recent considerations have implied $\theta_{13}\neq 0$
\cite{A41, A42, A43, A44, A45,A46, A47, A48, A49, A410, A411, A412,
A413, A414, A415, A416, A417, A418, S31,S32,S33,S34,S35,S36,S37,
S38,S39,S310,S311,S312,S313,S314,
S315,S316,S317,S318,S319,S320,S321,S322,S323,S324,S325,S326,
S327,S328,S329,S330,S331,S332,S333,S334,S335,S336,S337,S338,S339,S340,S341,S342, S41,S42,S43,S44,S45,S46,S47,S48,S49,S410,S411,S412,S413,S414,
S415,S416,S417,S418,S419,S420,S421,S422,S423,S424,S425,S426,S427,S428,S429} ,
but small as given in (\ref{PDG2012}).
     This problem has been improved in Ref. \citen{dlnvS3} by adding a new triplet $\rho$
     and another antisextet $s'$, in which $s'$ is regarded as a small perturbation.
     Therefore the model contain up to eight Higgs multiplets, and the scalar potential
of the model is quite complicated.

In this paper, we investigate
another choice with $D_4$, the group of a square, which is
 the second smallest non-Abelian
discrete symmetry. $D_4$ contains one doublet irreducible
representation and four singlets. This feature is useful to
separate the third family of fermions from the others. The group
contains a $\underline{2}$ irreducible representation which can
connect two maximally mixed generations. Besides the facilitating
maximal mixing through $\underline{2}$, it provides four
inequivalent singlet representations $\underline{1},
\underline{1}', \underline{1}''$ and $\underline{1}'''$ which play
a crucial role in consistently reproducing fermion masses and
mixing as a perturbation. We will  point out that this model is
simpler than the $S_3$ one, since fewer Higgs multiplets are needed in
order to allow the fermions to gain masses and to break symmetries and the
physics we will see is different from the former.
On the other hand, the neutrino sector is more simpler than that of $S_3$ one.
The boson masses and mixings are considered more generally and more detail
than those in Ref. \citen{DongHLT}.

There are two typical variants of the 3-3-1 models as far as lepton
sectors are concerned. In the minimal version, three
$\mathrm{SU}(3)_L$ lepton triplets are $(\nu_L,l_L,l^c_R)$, where
$l_{R}$ are ordinary right-handed charged-leptons \cite{331m1,
331m2, 331m3, 331m4, 331m5}. In
the second version, the third components of lepton triplets
are the  right-handed neutrinos,
$(\nu_L,l_L,\nu^c_R)$ \cite{331r1,331r2,331r4,331r5,331r6}   . To have a model with the
realistic neutrino mixing matrix, we should consider another
variant of the form $(\nu_L,l_L,N^c_R)$ where $N_R$ are three new
fermion singlets under standard model symmetry with vanishing
lepton-numbers \cite{dlshA4, dlsvS4}.

The rest of this work is organized
as follows. In Sec. \ref{fermion} and \ref{Chargedlep} we present
the necessary elements of the 3-3-1 model with the $D_4$ symmetry
as well as introducing necessary Higgs fields responsible for the
charged lepton masses. In Sec. \ref{quark}, we  discuss on quark
sector. Sec. \ref{neutrino} is devoted for the neutrino mass and
mixing. in Sec. \ref{vev} we consider the Higgs potential and minimization conditions.
Sec. \ref{Gaugeboson} is devoted for the gauge boson mass
and mixing. We summarize our results and make conclusions in the
section \ref{conclus}. Appendix \ref{apa} presents a brief of the
$D_4$ theory. Appendix \ref{apb} provides the lepton number ($L$)
and lepton parity ($P_l$) of  particles in the model.

\section{Fermion content\label{fermion}}

The gauge symmetry is based on $\mathrm{SU}(3)_C\otimes
\mathrm{SU}(3)_L \otimes \mathrm{U}(1)_X$, where the electroweak
factor $\mathrm{SU}(3)_L \otimes \mathrm{U}(1)_X$ is extended from
those of the SM where the strong interaction sector is retained.
Each lepton family includes a new electrically- and
leptonically-neutral fermion $(N_R)$ and is arranged under the
$\mathrm{SU}(3)_L$ symmetry as a triplet $(\nu_L, l_L, N^c_R)$ and
a singlet $l_R$. The residual electric charge operator $Q$ is
therefore related to the generators of the gauge symmetry by \[
Q=T_3-\fr{1}{\sqrt{3}}T_8+X, \] where $T_a$ $(a=1,2,...,8)$ are
$\mathrm{SU}(3)_L$ charges with $\mathrm{Tr}T_aT_b=\fr 1 2
\de_{ab}$ and $X$ is the  $\mathrm{U}(1)_X$ charge. This means
that the model under consideration does not contain exotic
electric charges in the fundamental fermion, scalar and adjoint
gauge boson representations.

Since the particles in the lepton triplet have different lepton
number (1 and 0), so the lepton number in the model  does not
commute with the gauge symmetry unlike the SM. Therefore, it is
better to work with a new conserved charge $\mathcal{L}$ commuting
with the gauge symmetry and related to the ordinary lepton number
by diagonal matrices \cite{dlshA4,dlsvS4, clong}
 \[L=\fr{2}{\sqrt{3}}T_8+\mathcal{L}.\]

The lepton charge arranged in this way (i.e. $L(N_R)=0$ as
assumed) is in order to prevent unwanted interactions due to
$\mathrm{U}(1)_\mathcal{L}$ symmetry and breaking  (due to the
lepton parity as shown below) to obtain the consistent lepton and
quark spectra. By this embedding, exotic quarks $U, D$ as well as
new non-Hermitian gauge bosons $X^0$, $Y^\pm$ possess lepton
charges as of the ordinary leptons:
$L(D)=-L(U)=L(X^0)=L(Y^{-})=1$.  A brief of the theory of $D_4$
group is given in Appendix \ref{apa}. The $D_4$ contains one
doublet irreducible representation $\underline{2}$ and four
singlets $\underline{1}$, $\underline{1}'$, $\underline{1}''$ and
$\underline{1}'''$. In this paper we work in real basic, in which
the two - dimensional representation $\underline{2}$ of $D_4$ is
real, $2^*(1^*, 2^*)=2(1^*, 2^*)$, and \bea \underline{2}(1,2)
\otimes \underline{2}(1,2) =\underline{1}(11+22) \oplus
\underline{1}'(11-22) \oplus \underline{1}{''}(12+21)\oplus
\underline{1}{'''}(12-21).\eea In the model under consideration,
we put the first family of  leptons in singlets $\underline{1}$ of
$D_4$, while the two other families are in the doublets
$\underline{2}$. Under the $[\mathrm{SU}(3)_L, \mathrm{U}(1)_X,
\mathrm{U}(1)_\mathcal{L},\underline{D}_4]$ symmetries as
proposed, the fermions of the model transform as follows \bea
\psi_{1L} &\equiv& \left( \nu_{1L} \hs
    l_{1L} \hs
    N^c_{1R}\right)^T\sim [3,-1/3,2/3,\underline{1}],\crn
\psi_{i L }&=&
\left(\nu_{i L}  \hs
    l_{i L}  \hs
    N^c_{i R} \right)^T\sim [3,-1/3,2/3,\underline{2}],\crn
l_{1R}&\sim&[1,-1,1,\underline{1}],\hs
l_{i R}\sim[1,-1,1,\underline{2}],\hs\hs\hs\,\, (i=2, 3)\crn
 Q_{3L}&=& \left(u_{3L} \hs
    d_{3L} \hs
    U_{L} \right)^T\sim[3,1/3,-1/3,\underline{1}],\crn
 Q_{\al L}&\equiv &
 \left( d_{\al L} \hs
  -u_{\al L}  \hs
    D_{\al L}\right)^T\sim[3^*,0,1/3,\underline{2}], \crn
u_{3R} &\sim &[1,2/3,0,\underline{1}], \hs u_{\al R}
\sim[1,2/3,0,\underline{2}]\crn
 d_{3R}&\sim&[1,-1/3,0,\underline{1}], \hs d_{\al  R}
 \sim[1,-1/3,0,\underline{2}],\crn
U_R&\sim&[1,2/3,-1,\underline{1}],\hs D_{\al R}
\sim[1,-1/3,1,\underline{2}], \hs (\al = 1, 2).\nn\eea where the
subscript numbers on field indicate to respective families which
also  in order define components of their $D_4$ multiplets. In the following, we consider
possibilities of generating the masses for the fermions. The
scalar multiplets needed for the purpose are also introduced.

\section{Charged lepton masses\label{Chargedlep}}
To generate masses for charged leptons, we need a minimum of
five $SU(3)_L$ Higgs triplets that lying in $\underline{1},
\underline{1}',\underline{1}'',\underline{1}'''$ and
$\underline{2}$. In decomposing of $\underline{2}\otimes
\underline{2}$ into irreducible representations, there is no
$\underline{2}$ one. So, it is required two Higgs scalars
\bea \phi = \left(%
\begin{array}{c}
  \phi^+_1 \\
  \phi^0_2 \\
  \phi^+_3 \\
\end{array}%
\right)\sim [3,2/3,-1/3, \underline{1}],\hs \phi{'} = \left(%
\begin{array}{c}
  \phi'^+_1 \\
  \phi'^0_2 \\
  \phi'^+_3 \\
\end{array}%
\right)\sim [3,2/3,-1/3, \underline{1}{'}], \label{phiphip}\eea
with VEVs as follows:
\bea \langle \phi \rangle = \left(  0, \hs v, \hs 0 \right)^T,\hs \langle \phi{'} \rangle = \left(
  0 \hs
  v' \hs
  0 \right)^T.\label{vevphiphip} \eea
The Yukawa interactions are
 \bea -\mathcal{L}_{l}&=&h_1 (\bar{\psi}_{1L}
\phi)_{\underline{1}} l_{1R}+h_2 (\bar{\psi}_{i L}\phi)_{\underline{2}}l_{i R}
+h_3 (\bar{\psi}_{i L}\phi{'})_{\underline{2}} l_{i R}+h.c\crn
&=&h_1 (\bar{\psi}_{1L}
\phi)_{\underline{1}} l_{1R}
+h_2\left(\bar{\psi}_{2 L}\phi l_{2 R}+\bar{\psi}_{3 L}\phi l_{3 R}\right)
 +h_3\left(\bar{\psi}_{2 L}\phi' l_{2 R}-
\bar{\psi}_{3 L}\phi' l_{3 R}\right)+h.c.\nn \eea
The mass Lagrangian of the
charged leptons reads \[
-\mathcal{L}^{\mathrm{mass}}_l=(\bar{l}_{1L},\bar{l}_{2L},\bar{l}_{3L})
M_l (l_{1R},l_{2R},l_{3R})^T+h.c.\]
 \bea M_l=
\left(%
\begin{array}{ccc}
  h_1v & 0 & 0 \\
   0 & h_2v+h_3v'& 0\\
  0 &  0 & h_2v-h_3v'\\
\end{array}%
\right)=
\left(%
\begin{array}{ccc}
  m_e & 0 & 0 \\
   0 & m_{\mu}& 0\\
  0 &  0 & m_{\tau}\\
\end{array}%
\right).\nn \eea
It is then diagonalized, and
\[ U_{eL}^+= U_{eR}=I \]
This means that the charged leptons $l_{1,2,3}$ by themselves are the physical mass eigenstates,
and the lepton mixing matrix depends on only that
of the neutrinos that will be studied in  section \ref{neutrino}.

 We see that the masses of muon and tauon are
 separated by the $\phi{'}$ triplet. This is the reason
why we introduce $\phi{'}$ in addition to $\phi$. The charged
lepton Yukawa couplings $h_{1, 2, 3}$ relate to their masses
as follows:
\bea h_1 v &=& m_e,\crn 2 h_2 v &=& m_\tau + m_\mu,\label{htkll}\\
 2 h_3 v' &=&  m_\mu-m_\tau. \nn
 \eea
The experimental  values for masses of  the charged leptons at the weak
scale are given as \cite{PDG2010, PDG20101}  : \bea m_e=0.511\
\textrm{MeV},\hs \ m_{\mu}=106.0 \ \textrm{MeV},\hs m_{\tau}=1.77\
\textrm{GeV} \eea Thus, we get \bea h_1 v=0.511\ \textrm{MeV},\
h_2v =938\ \textrm{MeV},\   |h_3 v'|=832 \ \textrm{MeV}\label{tn}
\eea It follows that if $v'$ and $v$ are of the same order of
magnitude, $h_1\ll h_2$ and $h_2\sim |h_3|$ .

\section{\label{quark}Quark masses}
\hs To generate masses for quarks with a minimal Higgs content, we
additionally introduce the following Higgs triplets \bea \chi&=&
\left(  \chi^0_1 \hs
  \chi^-_2 \hs
  \chi^0_3 \right)^T\sim[3,-1/3,2/3,\underline{1}],\crn
\eta&=&
\left( \eta^0_1 \hs
  \eta^-_2 \hs
  \eta^0_3\right)^T\sim[3,-1/3,-1/3,\underline{1}],\crn
  \eta{'}&=&
\left( \eta'^0_1 \hs
  \eta'^-_2 \hs
  \eta'^0_3 \right)^T\sim[3,-1/3,-1/3,\underline{1}{'}].\label{chietaetap}\eea
The Yukawa interactions are: \bea -\mathcal{L}_q &=&
f_3(\bar{Q}_{3L}U_R)_{\underline{1}}\chi + f (\bar{Q}_{\al
L}D_{\al R})_{\underline{1}}\chi^*\crn
&+&h^d_{3}(\bar{Q}_{3L}d_{3R})_{\underline{1}}\phi +
h^d(\bar{Q}_{\alpha L}d_{\alpha R})_{\underline{1}}\eta^*
+h'^d(\bar{Q}_{\alpha L}d_{\alpha
R})_{\underline{1}'}\eta{'}^*\crn &+&h^u_{3}
(\bar{Q}_{3L}u_{3R})_{\underline{1}}\eta + h^u (\bar{Q}_{\alpha
L}u_{\alpha R})_{\underline{1}}\phi^* + h'^u(\bar{Q}_{\alpha
L}u_{\alpha R})_{\underline{1}'}\phi{'}^*+h.c.\crn &=&
f_3(\bar{Q}_{3L}U_R)_{\underline{1}}\chi + f\left( \bar{Q}_{1
L}D_{1 R}+\bar{Q}_{2 L}D_{2 R}\right) \chi^*\crn
&+&h^d_{3}(\bar{Q}_{3L}d_{3R})_{\underline{1}}\phi +h^d\left(
\bar{Q}_{1 L}d_{1 R} +\bar{Q}_{2 L} d_{2 R}\right)\eta^* +h'^d
\left(\bar{Q}_{1 L}d_{1 R}-\bar{Q}_{2 L}d_{2
R}\right)\eta{'}^*\crn &+&h^u_{3}
(\bar{Q}_{3L}u_{3R})_{\underline{1}}\eta + h^u\left( \bar{Q}_{1
L}u_{1 R}+  \bar{Q}_{2 L}u_{2 R}\right)\phi^* +
h'^u\left(\bar{Q}_{1 L}u_{1 R}- \bar{Q}_{2 L}u_{2
R}\right)\phi{'}^*\crn &+&h.c\label{Lquarks}\eea
We now
introduce a residual symmetry of lepton number $P_l\equiv (-1)^L$,
called ``lepton parity'' \cite{dlnvS3, dlshA4}, in order to
suppress the mixing between ordinary quarks and exotic quarks (for
 lepton number of the model particles, see Appendix \ref{apb}).
The particles with even parity ($P_l=1$) have $L=0,\pm 2$ and the
 particles with odd parity $(P_l=-1)$ have  $L=\pm 1$. In this
framework we assume that the lepton parity is an exact symmetry,
not spontaneously broken. This means that  due to the lepton
parity conservation, the fields carrying  lepton number ($L=\pm
1$) $\eta_3$, $\eta'_3$ and $\chi_1$ cannot develop VEV.
Suppose that the VEVs of $\chi$,
$\eta$ and $\eta{'}$ are \bea \langle\chi\rangle&=&
\left(%
\begin{array}{c}
  0 \\
  0 \\
  \om \\
\end{array}%
\right),\,
\langle\eta\rangle=
\left(%
\begin{array}{c}
  u \\
  0 \\
  0 \\
\end{array}%
\right),\hs
 \langle\eta{'}\rangle=
\left(%
\begin{array}{c}
  u' \\
  0 \\
  0 \\
\end{array}%
\right),\label{vevchietaetap}\eea then the exotic quarks  get
masses \bea m_U=f_3 w,\hs m_{D_{1,2}}=f w, \eea and the mass
Lagrangian of the ordinary quarks reads: \bea
-\mathcal{L}^{mass}_q &=&h^d_{3}v\bar{d}_{3L}d_{3R}+h^du\left(\bar{d}_{1
L}d_{1 R} +\bar{d}_{2 L} d_{2 R}\right) +h'^d
u'\left(\bar{d}_{1 L}d_{1 R}-\bar{d}_{2 L}d_{2 R}\right)\crn
&+&h^u_{3}u \bar{u}_{3L}u_{3R} - h^uv\left( \bar{u}_{1 L}u_{1 R}+
\bar{u}_{2 L}u_{2 R}\right) -h'^uv'\left(\bar{u}_{1 L}u_{1
R}-\bar{u}_{2 L}u_{2 R}\right) +h.c\crn
&=&(\bar{u}_{1L},\bar{u}_{2L},\bar{u}_{3L}) M_u (u_{1R}, u_{2R},
u_{3R})^T+(\bar{d}_{1L},\bar{d}_{2L},\bar{d}_{3L}) M_d (d_{1R},
d_{2R}, d_{3R})^T+h.c.\crn
 \label{LquarkD4exp}\eea From
(\ref{LquarkD4exp}), the mass matrices for the ordinary up-quarks
and down-quarks are, respectively, obtained as follows:\bea M_u&=&
\left(%
\begin{array}{ccc}
  -h^u v-h'^uv' & 0 & 0 \\
  0 & -h^u v+h'^uv'& 0 \\
  0 & 0 & h^u_{3}u \\
\end{array}%
\right)=\left(%
\begin{array}{ccc}
  m_u & 0 & 0 \\
  0 & m_c & 0 \\
  0 & 0 & m_t \\
\end{array}%
\right), \crn M_d&=&
\left(%
\begin{array}{ccc}
  h^d u+h'^d u' & 0 & 0 \\
  0 & h^du-h'^du' & 0 \\
  0 & 0 & h^d_{3}v \\
\end{array}%
\right)
=\left(%
\begin{array}{ccc}
  m_d & 0 & 0 \\
  0 & m_s & 0 \\
  0 & 0 & m_b \\
\end{array}%
\right).\label{quarkmasses}\eea
In similarity to the charged
leptons, the  masses of  $u$ and $c$ quarks are also separated by
the $\phi{'}$ scalar. We see also that the introduction of
$\eta{'}$ in addition to $\eta$ is necessary to provide the
different masses for $d$ and $s$ quarks. The expression (\ref{quarkmasses}) leads to the
relations:
 \bea h^u_{3}u & = &m_t,\,-2h^uv=m_u+m_c ,\,-2h'^uv'=m_u-m_c,\crn
 h^d_{3}v & = & m_b,\,2h^du=m_d+m_s ,\,2h'^du'=m_d-m_s. \nn \eea
The current mass values for the quarks are given by \cite{PDG2010, PDG20101}
\bea m_u&=&(1.8\div3.0)\ \textrm{MeV},\hs\,\,\, m_d=(4.5\div 5.5)\
\textrm{MeV},\hs\,\,\,\,\,\, m_c=(1.25\div1.30)\ \textrm{GeV},\crn
m_s&=&(90.0\div100.0)\ \textrm{MeV},\,\, m_t=(172.1\div 174.9)\
\textrm{GeV},\,\, m_b=(4.13\div4.37)\
\textrm{GeV}.\label{vien3}\crn\eea
Hence
 \bea h^u_{3}u & = &
(172.1\div 174.9)\ \textrm{GeV},\hs h^d_{3}v = (4.13\div4.37)\
\textrm{GeV},\crn
 |h^uv|&=& (625.9 \div 651.5) \ \textrm{MeV},\hs
h^du = (47.25 \div 52.75)\, \textrm{MeV},\crn
 |h'^d u'|&=& (42.75
\div 47.25)\, \textrm{MeV}\hs
 h'^uv' =  (624.1 \div 648.5)\, \textrm{MeV}.  \label{vien3}\eea
It is obvious that if $|u| \sim|v| \sim |v'| \sim |u'|$, the
Yukawa coupling hierarchies are $|h'^d| \sim h^d\ll h^u\sim h'^u \ll h^u_3, h^d_3$,
and the couplings between up-quarks
$(h^u, h'^u, h^u_3)$ and Higgs scalar multiplets are slightly
heavier than those of down-quarks $(h^d, h'^d, h^d_3)$,
respectively.

The unitary matrices, which couple the left-handed up and
down-quarks to those in the mass bases, are $U^u_L=1$ and
$U^d_L=1$, respectively. Therefore we get the CKM matrix \be
U_\mathrm{CKM}=U^{d\dagger}_L U^u_L=1.\label{a41}\ee This is a
good approximation for the realistic quark mixing matrix, which
implies that the mixings among the quarks are dynamically small.
The small permutations such as a breaking of the lepton parity due
to the odd VEVs $\langle \eta^0_3\rangle$, $\langle
\eta'^0_3\rangle$, $\langle\chi^0_1\rangle$, or a violation of
$\mathcal{L}$ and/nor $D_4$ symmetry due to unnormal Yukawa
interactions, namely $\bar{Q}_{3L}\chi u_{3R}$, $\bar{Q}_{\al L}
\chi^* d_{\al R}$, $\bar{Q}_{3L}\chi u_{\al R}$, $\bar{Q}_{\al
L}\chi^* d_{3 R}$ and so forth, will disturb the tree level matrix
resulting in mixing between ordinary and exotic quarks and
possibly providing the desirable quark mixing pattern. This also
leads to the flavor changing neutral current at the tree level but
strongly suppressed \cite{dlsvS4,dlshA4} .

Note that  $\bar{Q}_{\alpha L}d_{\alpha R}$ and
$\bar{Q}_{\alpha L}u_{\alpha R}$ transform as $1\oplus1'\oplus1''\oplus1'''$
under $D_4$. All terms of the Yukawa interactions responsible for quarks
 masses in (\ref{Lquarks}) are invariant  under the
  $[\mathrm{SU}(3)_L, \mathrm{U}(1)_X,
\mathrm{U}(1)_\mathcal{L},\underline{D}_4]$ symmetries.
If $\bar{Q}_{\alpha L}d_{\alpha R}$ and $\bar{Q}_{\alpha L}u_{\alpha R}$
 lying in $1''$ or/and $1'''$, the 1-2 mixing of ordinary quarks will take place.
 In this work, we add soft terms which violate
$D_4$ symmetry with $1''$. Hence, the total Lagrangian of the ordinary
 quarks is added two extra terms $ -{\Delta \mathcal{L}}^d_q$
 and $-{\Delta \mathcal{L}}^u_q$, given by
\bea -{\Delta \mathcal{L}}^d_q &=& k^d(\bar{Q}_{\alpha L}d_{\alpha R})_{\underline{1}''}\eta^*
+h.c.\crn
&=&k^du\bar{d}_{1 L}d_{2 R}+k^du\bar{d}_{2 L}d_{1 R}+h.c,\label{deltaqd}\\
 -{\Delta \mathcal{L}}^u_q &=&
k^u(\bar{Q}_{\alpha L}u_{\alpha R})_{\underline{1}''}\phi^*\crn
&=&-k^u v\bar{u}_{1 L}u_{2 R}-k^u v\bar{u}_{2 L}d_{1 R}+h.c.\label{deltaqu}\eea
 The total mass matrices for the  ordinary up-quarks
and down-quarks then take the form:
\bea M'_u&=&M_u+\Delta M_u=
\left(%
\begin{array}{ccc}
  -h^u v-h'^uv' & -k^uv & 0 \\
  -k^uv & -h^u v+h'^uv'& 0 \\
  0 & 0 & h^u_{3}u \\
\end{array}%
\right),\label{Mutotal}\\
M'_d&=&M_d+\Delta M_d=
\left(%
\begin{array}{ccc}
  h^du+h'^d u' & k^du & 0 \\
  k^du & h^du-h'^du' & 0 \\
  0 & 0 & h^d_{3}v \\
\end{array}%
\right).\label{Mqmix}\eea The $M'_u$ in (\ref{Mutotal}) is
diagonalized as
\[
V^{u+}_LM'_uV^u_R=\mathrm{diag}(m'_u,\, m'_c,\, m'_t),
\]
where
\bea
m'_u&=&\frac{[(k^u)^2-(h^u)^2]v^2+(h'^u)^2v'^2}
{\sqrt{(k^u)^2v^2+(h^uv-h'^uv')^2}},\crn
m'_c&=&\frac{[(k^u)^2-(h^u)^2]v^2+(h'^u)^2v'^2}
{\sqrt{(k^u)^2v^2+(h^uv+h'^uv')^2}},\hs m'_t \equiv m_t =h^u_3u.\label{mup}\eea
and
\[
V^u_R=1,\,\,
V^u_L=
\left(%
\begin{array}{ccc}
 \frac{h^uv-h'^uv'}
{\sqrt{(k^u)^2v^2+(h^uv-h'^uv')^2}}&\hs-\frac{k^uv}
{\sqrt{(k^u)^2v^2+(h^uv+h'^uv')^2}}&\,0 \\
-\frac{k^u v}
{\sqrt{(k^u)^2v^2+(h^uv+h'^uv')^2}}&\hs\frac{h^uv+h'^uv'}
{\sqrt{(k^u)^2v^2+(h^uv+h'^uv')^2}}&\,0 \\
  0 &\hs 0 & \,1 \\
\end{array}%
\right).\]
The terms in (\ref{deltaqu}) and (\ref{deltaqd}) violate the $D_4$ symmetry,
therefore they should be much weaker than those of in (\ref{Lquarks}). This means
that
\bea
k^u&\ll& h^u,\, h'^u,\label{ku}\\
k^d&\ll& h^d,\, h'^d.\label{kd}
\eea
From condition (\ref{ku}), it follows that  \[
\mathcal{P} \equiv \frac{2h^uk^uv^2}{\sqrt{(k^u)^2v^2+(h^uv-h'^uv')^2}
\sqrt{(k^u)^2v^2+(h^uv+h'^uv')^2}}\] is very small, and \[
V^{u+}_LV^u_L=\left(%
\begin{array}{ccc}
 1&-\mathcal{P}&0 \\
-\mathcal{P}&1&0 \\
 0& 0 & \,1 \\
\end{array}%
\right)\simeq I.\] Similarly, the $M'_d$ in (\ref{Mqmix}) is
diagonalized as
\[
V^{d+}_LM'_dV^d_R=\mathrm{diag}(m'_d,\, m'_s,\, m'_b),
\]
where
\bea
m'_d&=&\frac{[(k^d)^2-(h^d)^2]u^2+(h'^d)^2u'^2}
{\sqrt{(k^d)^2u^2+(h^du-h'^du')^2}},\crn
m'_s&=&\frac{[(k^d)^2-(h^d)^2]u^2+(h'^d)^2u'^2}
{\sqrt{(k^d)^2u^2+(h^du+h'^du')^2}},\hs m'_b\equiv m_b =h^d_3v,\label{mdp}\eea
and
\be
V^d_R=1,\,\,
V^d_L=
\left(%
\begin{array}{ccc}
 \frac{-(h^du-h'^du')}
{\sqrt{(k^d)^2u^2+(h^du-h'^du')^2}}&\hs-\frac{k^du}
{\sqrt{(k^d)^2u^2+(h^du+h'^du')^2}}&\,0 \\
\frac{k^du}
{\sqrt{(k^d)^2u^2+(h^du-h'^du')^2}}&\hs\frac{h^du+h'^du'}
{\sqrt{(k^d)^2u^2+(h^du+h'^du')^2}}&\,0 \\
  0 &\hs 0 & \,1 \\
\end{array}%
\right).\label{VLd}
\ee
Analogously, from condition (\ref{kd}), we see that the value defined as
\[
\mathcal{K} \equiv \frac{2h^dk^du^2}{\sqrt{(k^d)^2u^2+(h^du-h'^du')^2}
\sqrt{(k^d)^2u^2+(h^du+h'^du')^2}}\] is very small, and \[
V^{d+}_LV^d_L=\left(%
\begin{array}{ccc}
 1&\mathcal{K}&0 \\
\mathcal{K}&1&0 \\
 0& 0 & \,1 \\
\end{array}%
\right)\simeq I.\]
The CKM matrix is then takes the form:
\be
V_{CKM}=V^{u+}_LV^d_L=\left(%
\begin{array}{ccc}
 V_{11}&V_{12}&0 \\
V_{21}&V_{22}&0 \\
 0& 0 & \,1 \\
\end{array}%
\right),\label{Vckm}
\ee
where
\bea
V_{11}&=&\frac{-(h^uh^d+k^uk^d)uv+h^uh'^du'v+h'^uh^duv'-h'^uh'^du'v'}
{\sqrt{(k^d)^2u^2+(h^du-h'^du')^2}\sqrt{(k^u)^2v^2+(h^uv-h'^uv')^2}},\crn
V_{12}&=&\frac{-(h^uk^d+k^uh^d)uv-k^uh'^du'v+h'^uk^duv'}
{\sqrt{(k^d)^2u^2+(h^du+h'^du')^2}\sqrt{(k^u)^2v^2+(h^uv-h'^uv')^2}},\crn
V_{21}&=&\frac{(h^uk^d+k^uh^d)uv-k^uh'^du'v+h'^uk^duv'}
{\sqrt{(k^d)^2u^2+(h^du-h'^du')^2}\sqrt{(k^u)^2v^2+(h^uv+h'^uv')^2}},\crn
V_{22}&=&\frac{k^uk^duv+h^du(h^uv+h'^uv')+h'^du'(h^uv+h'^uv')}
{\sqrt{(k^d)^2u^2+(h^du+h'^du')^2}\sqrt{(k^u)^2v^2+(h^uv+h'^uv')^2}}.\nn
\eea
With the help of conditions (\ref{ku}) and (\ref{kd}) we have:
\[
V_{11} \simeq 1,\,V_{12}\simeq 0,\,
V_{21}\simeq 0,\,
V_{22}\simeq 1,
\]
and the $V_{CKM}$ in (\ref{Vckm}) becomes
\[
V_{CKM} \simeq I. \]
If $SU(3)_L$ Higgs triplet $\phi$ in (\ref{phiphip}) lying in
$\underline{2}$ under $D_4$, the $1-3$ and $2-3$ mixings of the
 ordinary quarks will take place. A detailed study on
these problems are out of the scope of this work and should be
skip.

 \section{\label{neutrino}Neutrino mass and mixing}

The neutrino masses arise from the couplings of $\bar{\psi}^c_{i
L} \psi_{i L}$, $\bar{\psi}^c_{1L} \psi_{1L}$ and
$\bar{\psi}^c_{1L} \psi_{i L}$ to scalars, where $\bar{\psi}^c_{i
L} \psi_{i L}$ transforms as $3^*\oplus 6$ under
$\mathrm{SU}(3)_L$ and $\underline{1}\oplus \underline{1}'\oplus
\underline{1}''\oplus \underline{1}'''$ under $D_4$;
 $\bar{\psi}^c_{1 L} \psi_{1 L}$ transforms as
$3^*\oplus 6$ under $\mathrm{SU}(3)_L$ and $\underline{1}$ under
$D_4$, and $\bar{\psi}^c_{1L} \psi_{i L}$ transforms as $3^*\oplus
6$ under $\mathrm{SU}(3)_L$ and $\underline{2}$ under $D_4$. For
the known scalar triplets $(\phi, \phi', \chi,\eta, \eta')$, the
available interactions are only $(\bar{\psi}^c_{i L} \psi_{i
L})\phi$ and $(\bar{\psi}^c_{i L} \psi_{i L})\phi'$, but
explicitly suppressed because of the $\mathcal{L}$-symmetry. We
will therefore propose new SU(3)$_L$ antisextets, lying in either $\underline{1}$, $\underline{1}'$,
$\underline{1}''$, or $\underline{1}'''$ under $D_4$,
interact with  $\bar{\psi}^c_{ L}\psi_{ L}$ to produce masses for the neutrino. To obtain a
realistic neutrino spectrum, the antisextets transform as follows
\bea \si&=&
\left(%
\begin{array}{ccc}
   \si^0_{11} &  \si^+_{12} &  \si^0_{13} \\
   \si^+_{12} &  \si^{++}_{22} &  \si^+_{23} \\
   \si^0_{13} &  \si^+_{23} &  \si^0_{33} \\
\end{array}%
\right)\sim [6^*,2/3,-4/3,\underline{1}], \crn
s_k &=&
\left(%
\begin{array}{ccc}
  s^0_{11} & s^+_{12} & s^0_{13} \\
  s^+_{12} & s^{++}_{22} & s^+_{23} \\
  s^0_{13} & s^+_{23} & s^0_{33} \\
\end{array}%
\right)_k \sim [6^*,2/3,-4/3,\underline{2}], \nn\eea where the
numbered subscripts on the component scalars are the
$\mathrm{SU}(3)_L$ indices, whereas $k=1,2$ is that of $D_4$. The
VEV of $s$ and $\si $ is set as $(\langle s_1\rangle,\langle s_2\rangle)$
under $D_4$, in which
 \bea
\langle \si \rangle&=&\left(%
\begin{array}{ccc}
  \la_{\si } & 0 & v_{\si } \\
  0 & 0 & 0 \\
  v_{\si } & 0 & \Lambda_{\si } \\
\end{array}%
\right),\label{vevsi}\\
\langle s_k\rangle&=&\left(%
\begin{array}{ccc}
  \la_{k} & 0 & v_{k} \\
  0 & 0 & 0 \\
  v_{k} & 0 & \Lambda_{k} \\
\end{array}%
\right). \label{vevs}\eea
Following the potential minimization
conditions, we have several VEV alignments. The first alignment is that
$\langle s_1\rangle=\langle s_2\rangle$ or $\langle s_1\rangle\neq
0=\langle s_2\rangle$ or $\langle s_1\rangle=0\neq \langle
s_2\rangle$ then the $D_4$ is broken into $Z_2$ that consists of the elements
 \{$e, a^3b$\} or \{$e, b$\} or \{$e, a^2b$\}, respectively. The
second one  is that $0\neq \langle s_1\rangle\neq\langle
s_2\rangle\neq0$, then the $D_4$ is broken into $\{\mathrm{identity}\}$ (or
$Z_2\rightarrow \{\mathrm{identity}\}$). In this
work, we impose the first case in the first  alignment of $D_4$
breaking, i.e., \[ \la_1=\la_2\equiv \la_s,\hs v_1=v_2\equiv
v_s,\hs \La_1=\La_2\equiv \La_s.\] And, we additionally introduce
another scalar triplet lying in either $\underline{1}', \underline{1}''$
or $\underline{1}'''$ responsible for breaking the $Z_2$ subgroup
as the second stage of $D_4$ breaking. This can be achieved by
introducing a new $\mathrm{SU}(3)_L$ triplet, $\rho$  lying in
$\underline{1}'''$ as follows
 \be \rho = \left(%
\begin{array}{c}
  \rho^+_1 \\
  \rho^0_2 \\
  \rho^+_3 \\
\end{array}%
\right)\sim [3,2/3,-4/3, \underline{1}'''], \label{rho}\ee with the VEV given
by \be
\langle \rho \rangle = (0,\,\,v_\rho,\,\,0)^T.\label{vevrho}\ee

 The Yukawa interactions are:
\bea -\mathcal{L}_\nu&=&\fr 1 2 x (\bar{\psi}^c_{1 L} \psi_{1 L})_{\underline{1}}\sigma
+\fr 1 2 y \left(\bar{\psi}^c_{2 L}\psi_{2 L}+ \bar{\psi}^c_{3 L}\psi_{3 L}\right)\sigma \crn
&+&\fr 1 2 z \left[(\bar{\psi}^c_{1 L} \psi_{i L}) s+(\bar{\psi}^c_{i L} \psi_{1 L}) s\right]
+\fr 1 2 \tau (\bar{\psi}^c_{i L} \psi_{i L})_{1'''}\rho +h.c\crn
&=&\fr 1 2 x (\bar{\psi}^c_{1 L} \psi_{1 L})_{\underline{1}}\sigma
+\fr 1 2 y \left(\bar{\psi}^c_{2 L}\psi_{2 L}+ \bar{\psi}^c_{3 L}\psi_{3 L}\right)\sigma \crn
&+&\fr 1 2 z \left(\bar{\psi}^c_{1 L} \psi_{2 L} s_1+ \bar{\psi}^c_{1 L} \psi_{3 L} s_2+
\bar{\psi}^c_{2 L} \psi_{1 L} s_1+ \bar{\psi}^c_{3 L} \psi_{1 L} s_2\right)\crn
&+&\fr 1 2\tau\left(v_{\rho}\bar{N}_{2 R}\nu_{3 L}-v_{\rho}\bar{\nu}^c_{2 L}N^c_{3 R} -
v_{\rho}\bar{N}_{3R}\nu_{2 L}+v_{\rho}\bar{\nu}^c_{3 L}N^c_{2 R}\right) +h.c.\label{matrix}
\eea
The neutrino mass Lagrangian  can be written in matrix form as follows
 \be -\mathcal{L}^{\mathrm{mass}}_\nu=\fr 1 2
\bar{\chi}^c_L M_\nu \chi_L+ h.c.,\label{nm}\ee where
\bea \chi_L&\equiv&
\left(\nu_L \hs
  N^c_R \right)^T,\hs\,\,\, M_\nu\equiv\left(%
\begin{array}{cc}
  M_L & M^T_D \\
  M_D & M_R \\
\end{array}%
\right), \crn
\nu_L&=&(\nu_{1L},\nu_{2L},\nu_{3L})^T,\,\,
N_R=(N_{1R},N_{2R},N_{3R})^T, \nn \eea
and the mass matrices are then obtained by
\be
M_{L,R,D}=\left(%
\begin{array}{ccc}
  a_{L,R,D} & b_{L,R,D} & b_{L,R,D} \\
  b_{L,R,D} & c_{L,R,D} & d_{L,R,D} \\
  b_{L,R,D} & -d_{L,R,D} & c_{L,R,D} \\
\end{array}%
\right),\label{abcd}\ee
with
\bea
  a_{L} & =&\la_\si x, \hs  a_{D} =v_\si x, \hs  a_{R} =\La_\si x, \crn
  b_{L} & =&\la_s z,\hs b_{D}= v_{s}z,\hs b_{R} =\La_{s}z,\crn
 c_{L} & = &\la_{\si}y,\hs c_{D}=v_{\si}y,\hs
c_{R}=\La_{\si }y,\crn
d_{L}&=&0,\hs d_D = v_\rho \tau\equiv d,\hs d_R=0. \label{bL,cL}\eea
Three observed neutrinos gain masses via a combination of type I
and type II seesaw mechanisms derived from (\ref{nm}) and (\ref{abcd}) as \be
M_{\mathrm{eff}}=M_L-M_D^TM_R^{-1}M_D=\left(%
\begin{array}{ccc}
  A & B_1 & B_2 \\
  B_1 & C_1 & D_1 \\
  B_2 & D_1 & C_2 \\
\end{array}%
\right), \label{Mef}\ee where \bea
A&=&\left(2a_Rb^2_D-4a_Db_Db_R+2a_Lb^2_R+a^2_Dc_R-a_La_Rc_R
\right)/\left(2b^2_R-a_Rc_R\right),\crn
B_1&=&\left[-2b^2_Db_R+b_L(2b^2_R-a_Rc_R)+b_D(a_Dc_R+a_Rc_D-a_Rd_D)
-a_Db_R(c_D-d_D)\right]\crn
&&/\left(2b^2_R-a_Rc_R\right),\crn
B_2&=&\left[-2b^2_Db_R+b_L(2b^2_R-a_Rc_R)+b_D(a_Dc_R+a_Rc_D
+a_Rd_D)-a_Db_R(c_D+d_D)\right]\crn
&&/\left(2b^2_R-a_Rc_R\right),\crn
C_1&=&\left[-2b_Db_Rc_R(c_D-d_D)-b^2_R[(c_D+d_D)^2-2c_Lc_R]
+c_R(b^2_Dc_R+a_Rc^2_D-a_Rc_Lc_R+a_Rd^2_D)\right]\crn
&&/[c_R(2b^2_R-a_Rc_R)],\crn
C_2&=&\left[-2b_Db_Rc_R(c_D+d_D)-b^2_R[(c_D-d_D)^2-2c_Lc_R]
+c_R(b^2_Dc_R+a_Rc^2_D-a_Rc_Lc_R+a_Rd^2_D)\right]\crn
&&/[c_R(2b^2_R-a_Rc_R)],\crn
D_1&=&\left[(b_Rc_D-b_Dc_R)^2-b^2_Rd^2_D\right]/[c_R(2b^2_R-a_Rc_R)].\label{AB12C12D}
\eea

\subsection{\label{norho}Experimental constraints in the case without the $\rho$  triplet}
 In the case without the $\rho$
contribution ($v_\rho=0$), $\la_1=\la_2=\la_s,\, v_1=v_2=v_s,\,
\La_1=\La_2=\La_s$, we have $B_1=B_2\equiv B,\ C_1 = C_2\equiv C,$
and $M_{\mathrm{eff}}$ in (\ref{Mef}) becomes \be
M^0_{\mathrm{eff}}=\left(\begin{array}{ccc}
  A & B & B \\
  B & C & D \\
  B & D & C \\
\end{array}%
\right),\label{M0nu}\ee where \bea
A&=&\frac{x\left[\La_\si (\La_\si \la_\si-v^2_\si)xy-2(\La^2_s\la_\si+\La_\si v^2_s-2
\La_s v_s v_\si)z^2\right]}{\La^2_{\si} xy-2\La^2_s z^2},\crn
B&=&\frac{\left[\La^2_\si \la_s+v_\si(\La_s v_\si-2\La_\si v_s)\right]xyz
+2\La_s(v^2_\si-\la_s\La_s)z^3}{\La^2_{\si} xy-2\La^2_s z^2},\crn
C&=&\frac{y\left\{\La^2_\si (\la_\si\La_\si-v^2_\si)xy-\left[\La^2_\si v^2_s-2\La_s
\La_\si v_s v_\si+\La^2_s(2\la_\si\La_s-v^2_\si)\right]z^2\right\}}{\La^3_{\si} xy-2\La^2_s \La_\si z^2},\crn
D&=&-\frac{(\La_\si v_s-\La_s v_\si)^2 yz^2}{\La^3_{\si} xy-2\La^2_s \La_\si z^2}
= -\frac{\La_s}{\La_\si} \frac{\La_syz^2}{\left(xy-2\frac{\La^2_s}{\La^2_\si} z^2\right)}
\left(\frac{v_s}{\La_s}-\frac{v_\si}{\La_\si}\right)^2\ll 1.\label{vienv1}\eea
This mass matrix takes the form similar to that of unbroken
$Z_2$ (i.e. $v_{\rho}=0$). However, the breaking of $Z_2$
($v_\rho\neq 0$) in this case is necessary to fit
the data (see below). Indeed, we can diagonalize $M^0_{\mathrm{eff}}$ in (\ref{M0nu}) as follows:
\[ U^T
M_{\mathrm{eff}}U=\mathrm{diag}(m_1,m_2,m_3),\]
 where \bea m_1&=&\fr 1
2 \left(A+C + D - \sqrt{8 B^2 + (A-C - D)^2}\right),\crn
 m_2&=&\fr 1 2 \left(A+C + D + \sqrt{8 B^2 + (A-C - D)^2}\right),\hs m_3=C-D,\label{m123}\eea
and the corresponding eigenstates put in the
lepton mixing matrix: \be U = \left(%
\begin{array}{ccc}
  \frac{K}{\sqrt{K^2+2}} & -\frac{\sqrt{2}}{\sqrt{K^2+2}}& 0 \\
  \frac{1}{\sqrt{K^2+2}} &\frac{1}{\sqrt{2}} \frac{K}{\sqrt{K^2+2}} & -\fr{1}{\sqrt{2}} \\
  \frac{1}{\sqrt{K^2+2}} &\frac{1}{\sqrt{2}} \frac{K}{\sqrt{K^2+2}} & \fr{1}{\sqrt{2}} \\
\end{array}%
\right)
= \left(%
\begin{array}{ccc}
  \frac{A-m_2}{\sqrt{(A-m_2)^2+2B^2}} & -\frac{\sqrt{2}B}{\sqrt{(A-m_2)^2+2B^2}}& 0 \\
  \frac{1}{\sqrt{2}}\frac{\sqrt{2}B}{\sqrt{(A-m_2)^2+2B^2}} &\frac{1}{\sqrt{2}}
  \frac{A-m_2}{\sqrt{(A-m_2)^2+2B^2}} & -\fr{1}{\sqrt{2}} \\
 \frac{1}{\sqrt{2}} \frac{\sqrt{2}B}{\sqrt{(A-m_2)^2+2B^2}} &\frac{1}{\sqrt{2}}
 \frac{A-m_2}{\sqrt{(A-m_2)^2+2B^2}} & \fr{1}{\sqrt{2}} \\
\end{array}%
\right),\label{U0}\ee with \[
K=\frac{A-C-D-\sqrt{8B^2+(A-C-D)^2}}{2B}.\]
Relations between $K$ and $m_1, m_2, m_3$ take the forms: \bea
&&m_1=\,\, KB+C+D,\, m_2=-KB+A,\, m_3=C-D,\crn
&&m_1+m_2+m_3=A+2C,\crn &&m_1m_2=-2B^2+A(C+D).
 \label{U}\eea
The $U$ matrix in (\ref{U0}) can be parameterized in three Euler's
angles, which implies: \be \theta_{13}=0,\hs
\theta_{23}=\pi/4,\hs \tan\theta_{12}= \fr{\sqrt{2}B}{A-m_2}\equiv
\fr{\sqrt{2}}{K}.\label{t12}\ee

The recent  data imply  that $\theta_{13}\neq 0$ \cite{PDG2012, PDG1, PDG2, PDG3, PDG4}.
If it is correct, this case will fail.  However,  the following
case improves this.

\subsection{Experimental constraints in the case with the $\rho$  triplet }

In this case with the $\rho$ contribution,
$v_\rho\neq 0$, the general neutrino mass matrix in (\ref{Mef}) can
be rewritten in the form:
\be M_{\mathrm{eff}}=\left(%
\begin{array}{ccc}
  A & B & B \\
  B & C & D \\
  B & D & C \\
\end{array}%
\right)+ \left(%
\begin{array}{ccc}
  \,\,\,0& \,\,\,p_1 & -p_1 \\
  \,\,\,\,p_1 & \,\,\,q_1 & \,\,\,r \\
  -p_1 & \,\,\,r & \,\,\,q_2 \\
\end{array}%
\right),\label{viensplit}\ee where $A, B, C$ and $D$ are given by
(\ref{vienv1})
due to the contribution from  the scalar antisextets $s$ and $\si$
only. The second  matrix in (\ref{viensplit}) is a deviation
arising from  the contribution due to the scalar triplet $\rho$,
namely $p_{1}=B_{1}- B=-(B_2-B)$, $q_{1,2}=C_{1,2} - C$ and
$r=D_1-D$, with the $A$, $B_{1,2}$, $C_{1,2}$ and $D_1$ being
defined in (\ref{AB12C12D}). Indeed, if the $\rho$ contribution is
neglected, the deviations $p_1, q_{1,2}, r$ will vanish. Hence the
mass matrix $M_{\mathrm{eff}}$ in (\ref{Mef}) reduces to its
 form in (\ref{M0nu}). The first term, as shown in subsection
\ref{norho} can approximately fit the Particle Data
Group 2010 \cite{PDG2010} with a small deviation for
$\theta_{13}$. The second term is proportional to $p_1,\ q_{1,2},\
r$ due to contribution of the triplet $\rho$, will take the role
for such a deviation of $\theta_{13}$. So, in this work we
consider the $\rho$ contribution as a small perturbation and
terminating the theory at the first order.

Assuming that $\la_s \ll v_s\ll\La_s, \ \la_{\si} \ll
v_{\si}\ll\La_{\si}$ or $\frac{\la_s}{\La_s} \simeq
\frac{\la_{\si}}{\La_{\si}} \ll 1,\, \frac{v_s}{\La_s} \simeq
\frac{v_{\si}}{\La_{\si}} \ll 1$, $v_{\rho}\ll v_s, v_{\si}$,
$\La_{\si}\sim \La_{s}$, and $x, y, z, \tau$ being in the same order
then we get \bea p_1 &=&
\frac{\tau v_{\rho}(\La_{\si}v_s-\La_sv_{\si})xz}{\La^2_{\si}xy-2\La^2_sz^2}
\simeq-\tau v_{\rho}\left(\frac{v_s}{\La_{s}}-
\frac{v_{\si}}{\La_{\si}}\right),\label{epsilonv1}\\
q_1&=&\frac{\tau v_{\rho}[-\La^2_{\si}\tau v_{\rho}xy
+\La_{s}(\La_{s}\tau v_{\rho}-2\La_{\si}v_sy+2\La_{s}v_{\si}y)z^2]}
{\La_{\si}y(\La^2_{\si}xy-2\La^2_{s}z^2)}\simeq2\tau v_{\rho}
\left(\frac{v_s}{\La_{s}}-\frac{v_{\si}}{\La_{\si}}\right),\label{q1}\\
q_2&=&\frac{\tau v_{\rho}[-\La^2_{\si}\tau v_{\rho}xy+\La_{s}
(\La_{s}tv_{\rho}+2\La_{\si}v_sy-2\La_{s}v_{\si}y)z^2]}
{\La_{\si}y(\La^2_{\si}xy-2\La^2_{s}z^2)}
\simeq-q_1,\label{q2}\eea
 which  all start from the first order of the perturbation
\be
r=\frac{\La^2_{s}\tau^2v^2_{\rho}z^2}{\La^3_{\si}xy^2-2\La^2_{s}\La_{\si}yz^2}
\simeq-\tau v_{\rho}\left(\frac{v_{\rho}}{\La_{\si}}\right).\label{r}\ee
Because of  $v_s\ll\La_s, \ v_{\si}\ll\La_{\si}$ and $v_{\rho}\ll
v_s, v_{\si}$ so $r$ in (\ref{r}) is the second order of the
perturbation. Consequently, it can be ignored.
 The last matrix in (\ref{viensplit}) now takes the form: \be
\left(%
\begin{array}{ccc}
  0& p_1 & -p_1 \\
  p_1 & q_1 & r \\
  -p_1 & r & q_2 \\
\end{array}%
\right)\approx \ep v_\rho \tau\left(%
\begin{array}{ccc}
  0& -1 & 1 \\
  -1 & 2 & 0 \\
  1 & 0 & -2 \\
\end{array}%
\right), \label{vien31}\ee
where $\ep=\frac{v_{s}}{\La_{s}}-\frac{v_{\si}}{\La_{\si}}$ is very small and
 plays the role of the perturbation  parameter.
 The explicit form of the mass matrix (\ref{Mef}) is thus given by
\be M_{\mathrm{eff}}=\left(%
\begin{array}{ccc}
  A & B & B \\
  B & C & D \\
  B & D & C \\
\end{array}%
\right)+ \ep M^{(1)},\,\label{Unuef1}\ee where $M^{(1)}$ is
 the perturbation contribution at the first order:
\[ M^{(1)} \equiv v_\rho \tau\left(%
\begin{array}{ccc}
  0& -1 & 1 \\
  -1 & 2 & 0 \\
  1 & 0 & -2 \\
\end{array}%
\right).\]
It is clear that the first term in
(\ref{Unuef1}) can approximately fit the data with a "small''
deviation as shown in subsection \ref{norho}. So, in this case we
consider $\epsilon$ being small as a perturbation parameter.

At the first order of perturbation theory, the matrix
$M^{(1)}$ does not give contribution to
  eigenvalues. However, it changes the eigenvectors. The  physical
  neutrino masses are thus obtained as:
\[ m'_{1} = m_1,\hs m'_{2}=m_2,\hs m'_3=m_3, \] where
$m_{1,2,3}$ are the masses in the case without contribution
of $\rho$ given by (\ref{m123}). For the corresponding perturbed
eigenstates, we put:
 \[ U\longrightarrow
U'= U+\Delta U,\] where $U$ is
 defined by (\ref{U0}), and
\be \Delta U= \left(%
\begin{array}{ccc}
0&\hs 0&\hs -\ep \frac{\sqrt{2}K(K-2)\tau v_\rho}{(K^2+2)(m_1-m_3)} \\
  -\ep \frac{(K-2)\tau v_\rho}{\sqrt{K^2+2}(m_1-m_3)}&\hs \,\,\,\,\ep
  \frac{\sqrt{2}(K+1)\tau v_\rho}{\sqrt{K^2+2}(m_2-m_3)}&\hs
  -\ep \frac{\sqrt{2}(K-2)\tau v_\rho}{(K^2+2)(m_1-m_3)} \\
 \,\,\,\,\ep \frac{(K-2)\tau v_\rho}{\sqrt{K^2+2}(m_1-m_3)}&\hs -
 \ep \frac{\sqrt{2}(K+1)\tau v_\rho}{\sqrt{K^2+2}(m_2-m_3)}&\hs
  -\ep \frac{\sqrt{2}(K-2)\tau v_\rho}{(K^2+2)(m_1-m_3)} \\
\end{array}%
\right).\label{detltaU}\ee  The lepton mixing matrix in this case
can still be parameterized in three new Euler's angles
$\theta'_{ij}$, which are also a perturbation from the
$\theta_{ij}$ (without contribution from the $\rho$ triplet),
defined by \bea s'_{13}&=&-U'_{13}=\ep
\frac{\sqrt{2}K(K-2)\tau v_\rho}{(K^2+2)(m_1-m_3)},\crn
 t'_{12}&=&-\frac{U'_{12}}{U'_{11}}
  =\frac{\sqrt{2}}{K}\equiv t_{12},\crn
  t'_{23}&=&-\frac{U'_{23}}{U'_{33}}=1+\fr{4\ep (K-2)v_\rho \tau}
  {2\ep(K-2) v_\rho \tau +( K^2+2)(m_1 - m_3)}.\nn
\eea It is easily to show that our model is consistent
since the five experimental constraints on the mixing angles and
squared mass differences of neutrinos can be respectively fitted
with four Yukawa coupling parameters $x,\ y,\ z$ and $\tau$
of the $s$, $\si$ antisextets and $\rho$ triplet scalars,
with the given VEVs.
To see this, let us take the data in 2012 as shown in (\ref{PDG2012}).
It follows $K\simeq 2.1054$, and
$t'_{23}=1.2383$ \, [$\theta'_{23}\simeq 51.08^o,
\,\sin^2(2\theta'_{23}) = 0.9556$ satisfying the condition
$\sin^2(2\theta'_{23})> 0.95$].

Until now values of neutrino masses (or the absolute neutrino
masses) as well as the mass ordering of neutrinos is unknown. The
tritium experiment \cite{Wein, Lobashev} provides an upper bound on the
absolute value of neutrino  mass \[ m_i\leq 2.2 \, \mathrm{eV} \]
A more stringent bound was found from the analysis of the latest
cosmological data \cite{Tegmark} \[ m_i\leq 0.6\,  \mathrm{eV}, \]
 while arguments from the growth of
large-scale structure in the early Universe yield the upper bound
\cite{Weiler} \[ \sum^{3}_{i=1} m_i\leq 0.5\,  \mathrm{eV}. \]

The neutrino mass spectrum can be the normal mass hierarchy ($
m_1\simeq m_2 < m_3$), the inverted hierarchy ($m_3< m_1\simeq m_2
$) or nearly degenerate ($m_1\simeq m_2\simeq m_3 $). The mass
ordering of neutrino depends on the sign of $\Delta m^2_{23}$
which is currently unknown. In the case of 3-neutrino mixing, in
the model under consideration,  the two possible signs of $\Delta
m^2_{23}$ correspond to two types of neutrino mass spectrum can be
provided as shown bellows.

\subsubsection{Normal case ($\Delta m^2_{23}> 0$)}
\begin{figure}[h]
\begin{center}
\includegraphics[width=8.0cm, height=4.0cm]{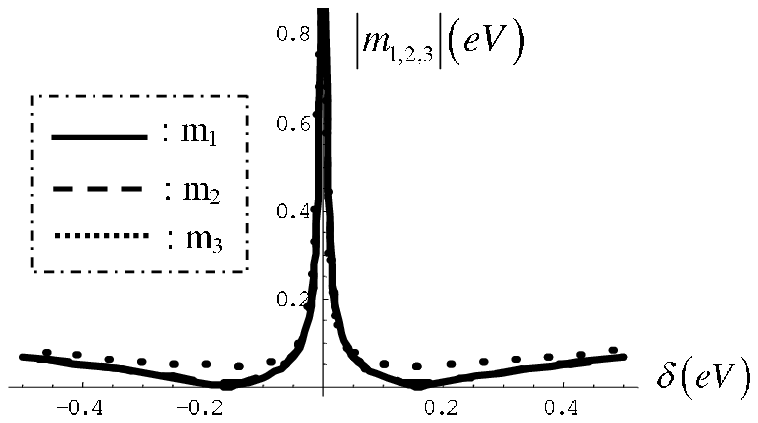}
\vspace*{-0.4cm} \caption[The absolute values of  $|m_1|, |m_2|,
|m_3|$ as  functions of $\delta$ with $\delta\in (-0.5, 0.5)\,
\mathrm{eV}$ in the case of $\Delta m^2_{23}> 0$]{The absolute
values $|m_1|, |m_2|, |m_3|$ as  functions of $\delta$ with
$\delta\in (-0.5, 0.5)\,  \mathrm{eV}$ in the case of $\Delta
m^2_{23}> 0$ }\label{m123w} \vspace*{0.5cm}

\includegraphics[width=12.2cm, height=5.0cm]{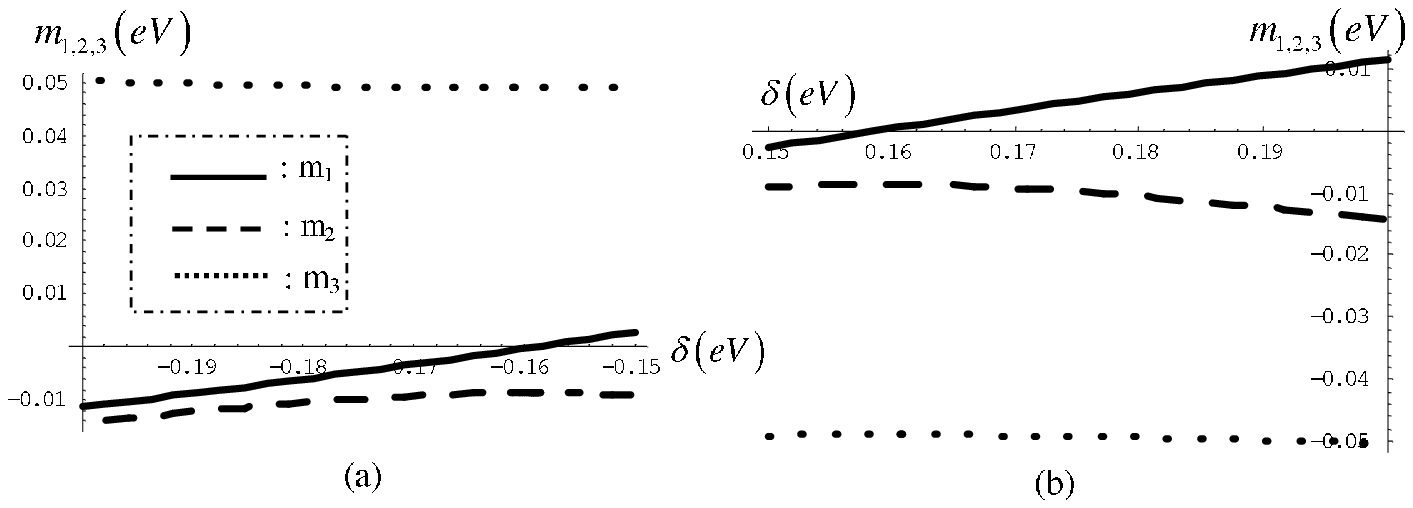}
\vspace*{-0.4cm} \caption[The values $m_i\, (i=1,2,3)$ as
functions of  $\delta$. (a) $\delta \in (-0.20, -0.15)\,
\mathrm{eV}$, (b) $\delta \in (0.15, 0.20) \, \mathrm{eV}$ in the
case of $\Delta m^2_{23}> 0$]{The  values
 $m_i\, (i=1,2,3)$ as  functions of  $\delta$. (a) $\delta \in (-0.20, -0.15)
\, \mathrm{eV}$, (b) $\delta \in (0.15, 0.20) \, \mathrm{eV}$ in
the case of $\Delta m^2_{23}> 0$}\label{m123d} \vspace*{0.5cm}

\includegraphics[width=12.0cm, height=5.0cm]{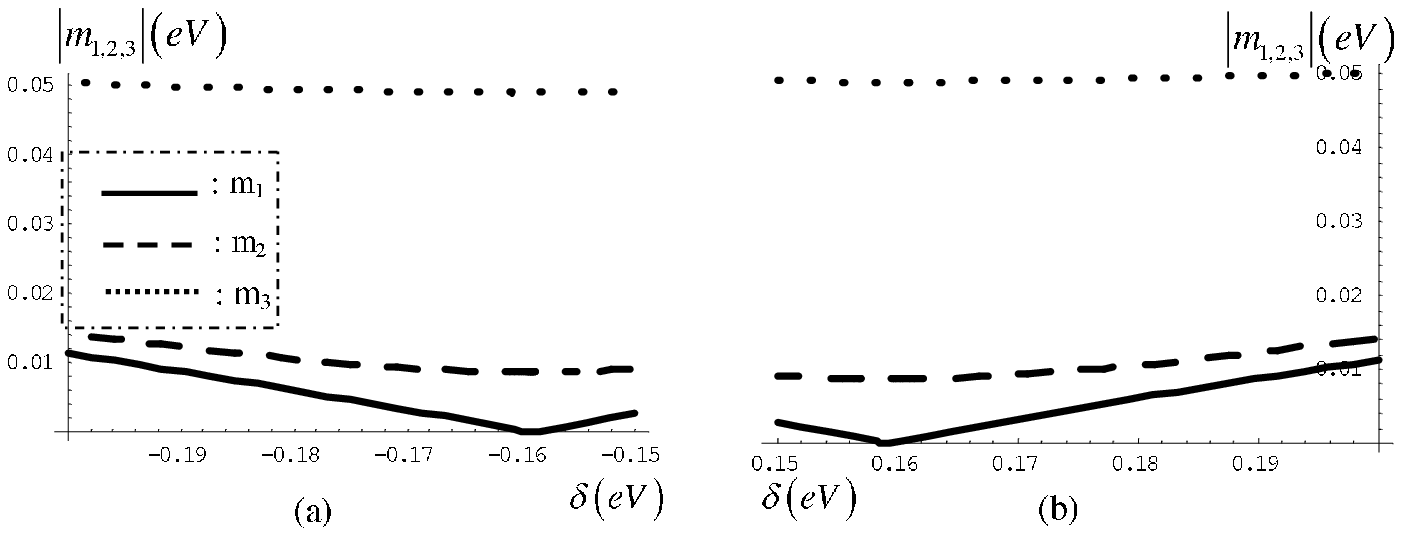}
\vspace*{-0.4cm} \caption[The absolute values  $|m_i|\, (i=1,2,3)$
as  functions of  $\delta$. (a) $\delta \in (-0.20, -0.15)\,
 \mathrm{eV}$, (b) $\delta \in (0.15, 0.20)\, \mathrm{eV}$ in the case of $\Delta m^2_{23}
 > 0$]{The absolute values $|m_i|\, (i=1,2,3)$ as  functions of  $\delta$. (a) $\delta \in (-0.20,
 -0.15)\, \mathrm{eV}$, (b) $\delta \in (0.15, 0.20) \,
 \mathrm{eV}$ in the case of $\Delta m^2_{23}> 0$}\label{m123a}
\vspace*{-0.3cm}
\end{center}
\end{figure}

\begin{figure}[h]
\begin{center}
\includegraphics[width=12.0cm, height=4.5cm]{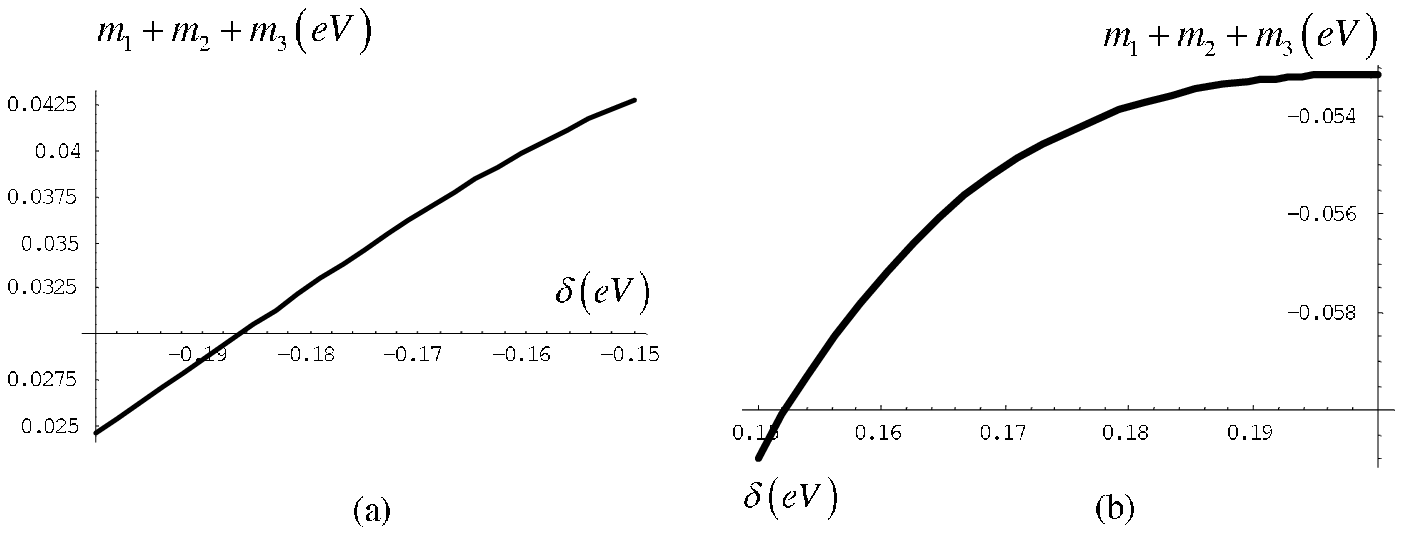}
\vspace*{-0.4cm} \caption[The value $\sum m_{i}$ as a function of
$\delta$ in the case of $\Delta m^2_{23}> 0$. (a) $\delta \in
(-0.20, -0.15)\, \mathrm{eV}$, (b) $\delta \in (0.15, 0.20)\,
\mathrm{eV}$]{The value $\sum m_{i}$ as a function of  $\delta$ in
the case of $\Delta m^2_{23}> 0$. (a) $\delta \in (-0.20, -0.15)\,
\mathrm{eV}$, (b) $\delta \in (0.15, 0.20)\,
\mathrm{eV}$}\label{m1m2m3s} \vspace*{0.5cm}

\includegraphics[width=12.0cm, height=4.5cm]{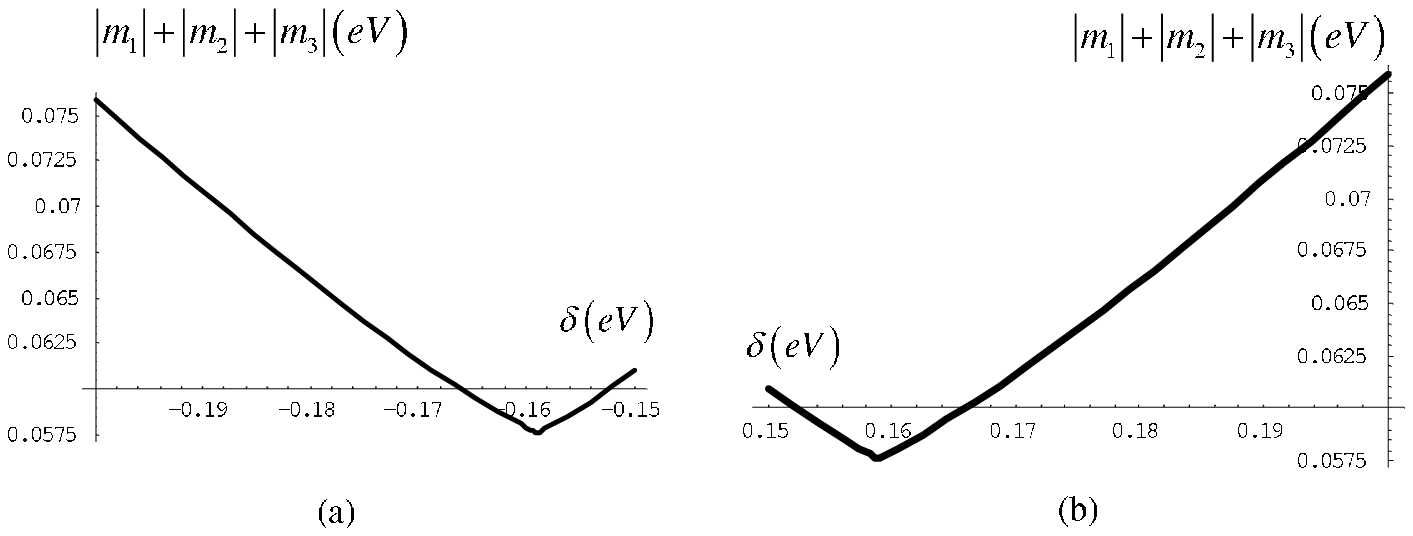}
\vspace*{-0.4cm} \caption[The value $\sum |m_{i}|$ as a function
of $\delta$ in the case of $\Delta m^2_{23}> 0$. (a) $\delta \in
(-0.20, -0.15)\, \mathrm{eV}$, (b) $\delta \in (0.15, 0.20)\,
\mathrm{eV}$]{The value $\sum |m_{i}|$ as a function of $\delta$ in
the case of $\Delta m^2_{23}> 0$. (a) $\delta \in (-0.20, -0.15)\,
\mathrm{eV}$, (b) $\delta \in (0.15, 0.20)\,
\mathrm{eV}$}\label{m1m2m3ada} \vspace*{-0.7cm}
\end{center}
\end{figure}

In this case, the neutrino masses are functions of $\delta=\ep
v_\rho \tau$ as follows
 \bea
&&m_1=-\frac{0.00388981}{\delta}+0.153928\, \delta,\label{m1p}\\
&&m_2=\pm 8\times 10^{-3}\sqrt{-17.5391+\frac{0.236416}{\delta^2}+370.216\,
\delta^2},\label{m2p}\\
&&m_3=-\frac{0.00388981}{\delta}-0.153928\, \delta,\label{m3p}\eea
In Fig. \ref{m123w}, we have plotted the absolute value $|m_i| \,
(i=1,2,3)$ as a function of $\delta$ with the values of $\delta\in
(-0.5, 0.5) \, \mathrm{eV}$. This figure shows that there exist
allowed regions for value of $\delta$ where
 either normal or quasi-degenerate neutrino masses spectrum achieved. The quasi-degenerate
  mass hierarchy obtained when $\delta\rightarrow 0$ or $\delta\rightarrow \pm \infty$ ($|\delta|$
  increase but must be  small enough because of the scale of $\ep, v_{\rho}, \tau$). The normal mass
   hierarchy will be  obtained if $\delta$ takes the values around $(-0.20,-0.15)\,
   \mathrm{eV}$ or (0.15, 0.20)
\,$\mathrm{eV}$ as shown in Figs. \ref{m123d}a and
\ref{m123d}b, respectively.
    The Figs. \ref{m123a}a and \ref{m123a}b
give three absolute neutrino masses $m_i$ with $\delta \in (-0.2,
-0.15)$ and $\delta \in (0.15, 0.20)$, respectively. The values
$\sum^3_{i=1}m_i$ as well as $\sum^3_{i=1}|m_i|$ as a functions of
$\delta$ are plotted in Figs. \ref{m1m2m3s} and \ref{m1m2m3ada},
respectively.
\\

  To get explicit values of the model
parameters, we assume $\delta \equiv \ep v_\rho \tau =0.15\, \mathrm{eV}$,  which is safely small.
 Then the neutrino masses are explicitly
 given as $m_1\simeq-0.00284\, \mathrm{eV}$, $m_2\simeq \pm0.00911\, \mathrm{eV}$
and $m_3\simeq-0.04902 \, \mathrm{eV}$. It follows that
 $A\simeq 8.75\times 10^{-4}\, \mathrm{eV}$, $B\simeq -3.913\times 10^{-3} \,\mathrm{eV}$,
 $C\simeq -2.18\times 10^{-2}\, \mathrm{eV}$ and
$D\simeq 2.72\times 10^{-2}\, \mathrm{eV}$ (equivalently to
$m_2=0.00911\, \mathrm{eV}$), or $A\simeq -7.16\times 10^{-3}\,
\mathrm{eV}$, $B\simeq -2.05\times 10^{-3} \,\mathrm{eV}$,
 $C\simeq -2.69\times 10^{-2}\, \mathrm{eV}$ and
$D\simeq 2.21\times 10^{-2}\, \mathrm{eV}$ (equivalently to
$m_2=-0.00911\, \mathrm{eV}$). This solution means a normal
neutrino mass spectrum as mentioned above. Furthermore, if
$\la_s=\la_{\si}=1\, \mathrm{eV},\, v_s=v_\si,\,
\La_s=-\La_\si=-v^2_s \, \mathrm{eV}$, we obtain $x\simeq
-1.57\times 10^{-4}$, $y \simeq -7.16\times 10^{-3}$, $z\simeq
1.09 \times 10^{-3}$ (equivalently to $m_2=0.00911\,
\mathrm{eV}$), or $x\simeq -3.68\times 10^{-3}$, $y \simeq -1.12
\times 10^{-2}$, $z\simeq -3.69 \times 10^{-2}$  (equivalently to
$m_2=-0.00911\, \mathrm{eV}$) and $\tau\simeq \frac{150}{v_\rho}$ .
If $v_\rho \sim1.5\times 10^{-3} \mathrm{GeV}$ then $\tau \sim1.5
\times 10^{-4}$ which is on the same order in magnitude with $x,
y, z$.

\subsubsection{Inverted case ($\Delta m^2_{23}< 0$)}
\begin{figure}[h]
\begin{center}
\includegraphics[width=8.0cm, height=4.0cm]{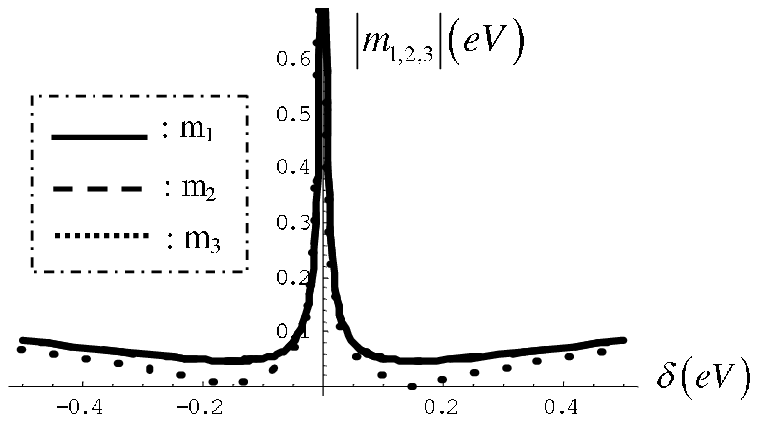}
\vspace*{-0.5cm} \caption[The absolute values $|m_1|, |m_2|,
|m_3|$ as  functions of  $\delta$ with $\delta\in (-0.5, 0.5)\,
\mathrm{eV}$ in the case of $\Delta m^2_{23}< 0$]{The absolute
values  $|m_1|, |m_2|, |m_3|$ as  functions of $\delta$ with
$\delta\in (-0.5, 0.5)\, \mathrm{eV}$ in the case of $\Delta
m^2_{23}< 0$ }\label{m123w1}

\includegraphics[width=12.0cm, height=4.5cm]{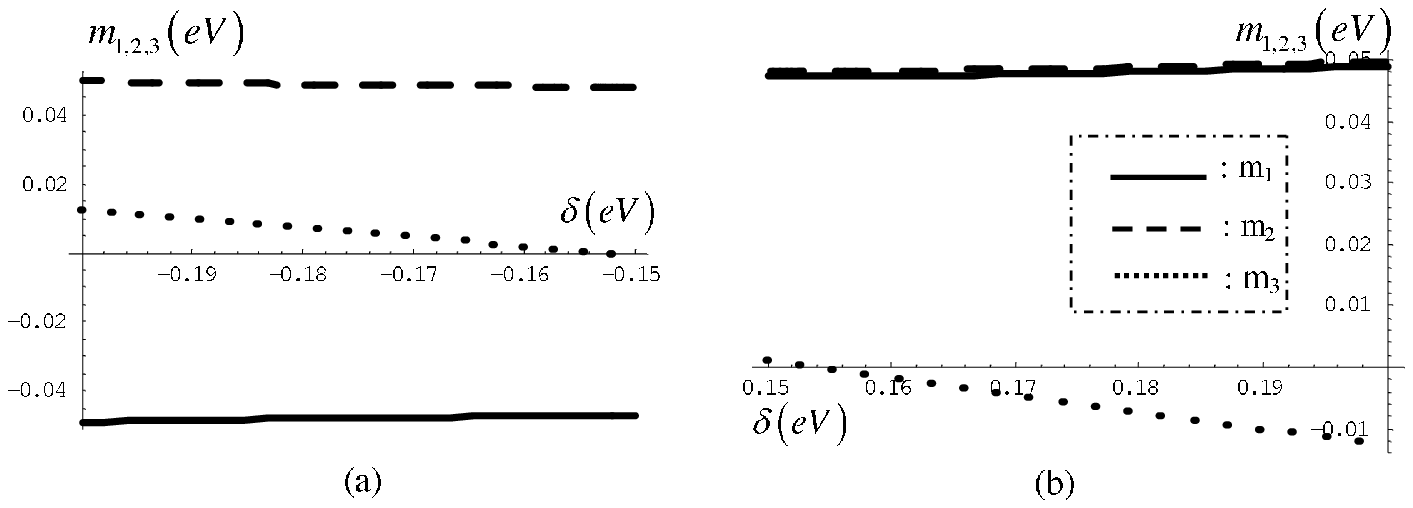}
\vspace*{-0.4cm} \caption[The value $m_i\, (i=1,2,3)$ as a
function of  $\delta$. (a) $\delta \in (-0.20, -0.15)\,
\mathrm{eV}$, (b) $\delta \in (0.15, 0.20)\, \mathrm{eV}$ in the
case of $\Delta m^2_{23}> 0$]{The values  $m_i\, (i=1,2,3)$ as
functions of  $\delta$. (a) $\delta \in (-0.20, -0.15)\,
\mathrm{eV}$, (b) $\delta \in (0.15, 0.20)\, \mathrm{eV}$ in the
case of $\Delta m^2_{23}< 0$}\label{m123d1}

\includegraphics[width=12.0cm, height=5.0cm]{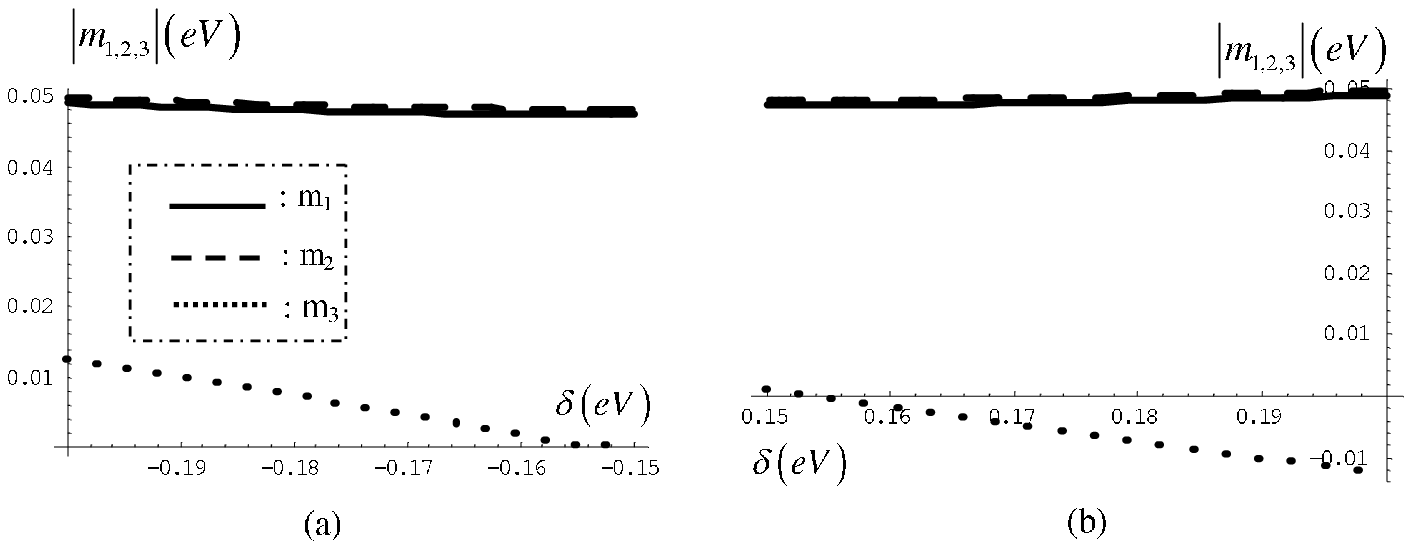}
\vspace*{-0.4cm} \caption[The absolute values $|m_i|\, (i=1,2,3)$
as  functions of $\delta$. (a) $\delta \in (-0.20, -0.15)
\mathrm{eV}$, (b) $\delta \in (0.15, 0.20) \mathrm{eV}$ in the
case of $\Delta m^2_{23}> 0$]{The absolute values $|m_i|\,
(i=1,2,3)$ as functions of  $\delta$. (a) $\delta \in (-0.20,
-0.15)\, \mathrm{eV}$, (b) $\delta \in (0.15, 0.20)\, \mathrm{eV}$
in the case of $\Delta m^2_{23}< 0$}\label{m123a1}
\end{center}
\end{figure}

\begin{figure}[h]
\begin{center}

\includegraphics[width=12.0cm, height=5.0cm]{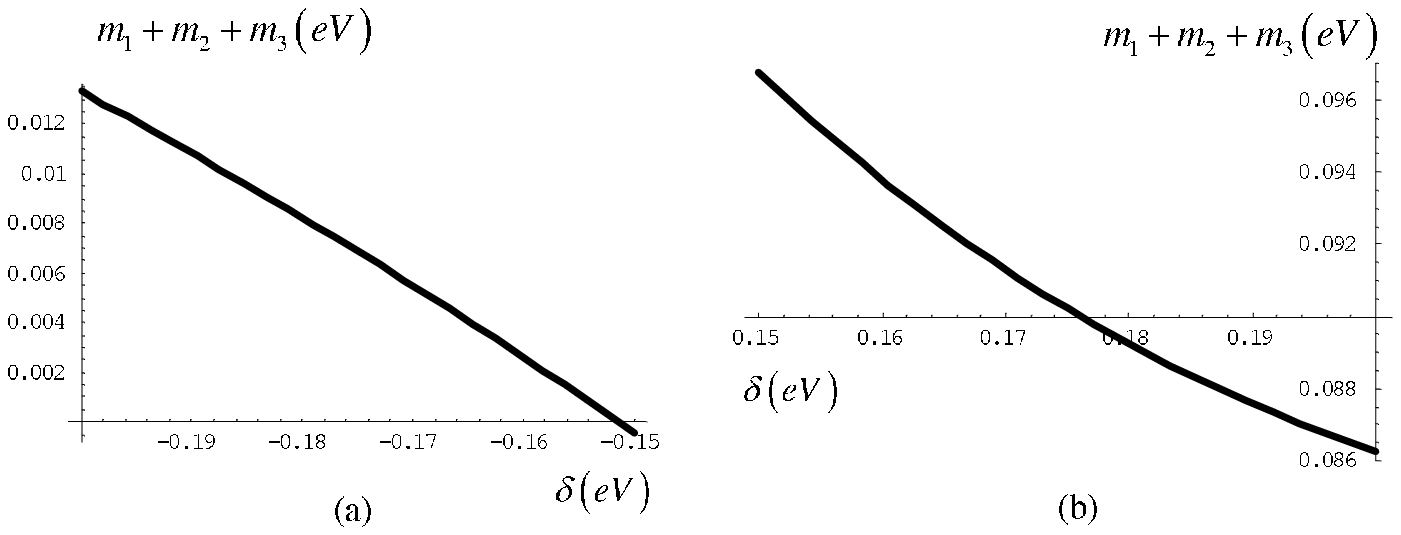}
\vspace*{-0.4cm} \caption[The value $\sum m_{i}$ as a function of
$\delta$ in the case of $\Delta m^2_{23}> 0$. (a) $\delta \in
(-0.20, -0.15)\, \mathrm{eV}$, (b) $\delta \in (0.15, 0.20)\,
\mathrm{eV}$]{The value $\sum m_{i}$ as a function of $\delta$ in
the case of $\Delta m^2_{23}< 0$. (a) $\delta \in (-0.20, -0.15)\,
\mathrm{eV}$, (b) $\delta \in (0.15, 0.20)\,
\mathrm{eV}$}\label{m1m2m3s1} \vspace*{0.8cm}
\end{center}
\end{figure}

\begin{figure}[h]
\begin{center}
\vspace*{-0.8cm}

\includegraphics[width=12.0cm, height=4.5cm]{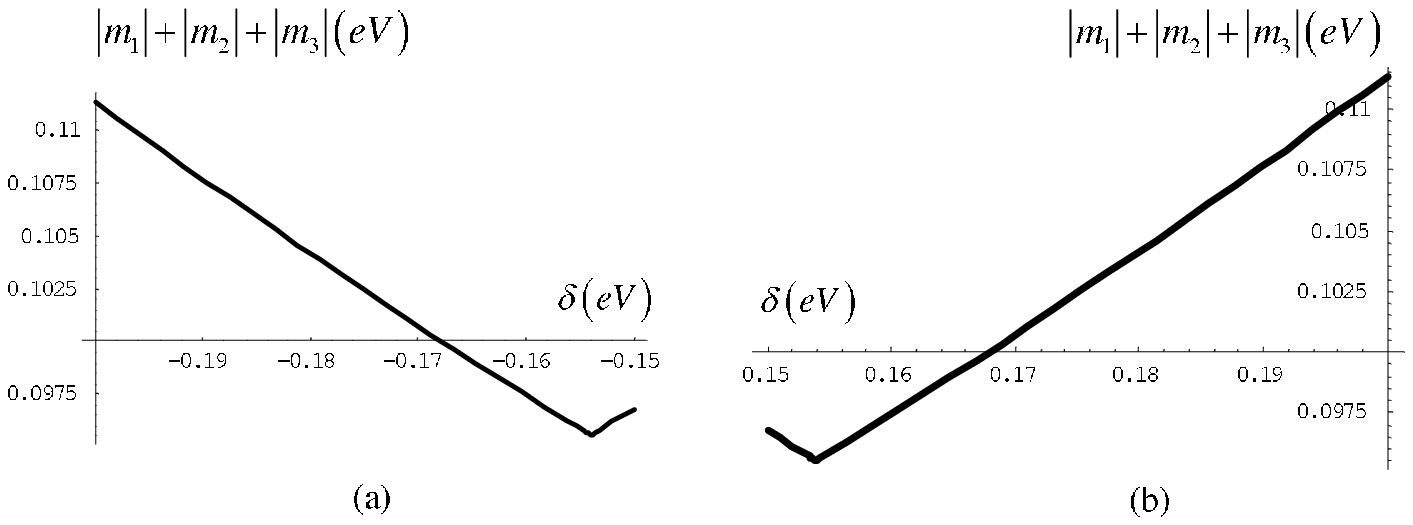}
\vspace*{-0.4cm}

 \caption[The values $m_{1,2,3}$ as functions of
$\delta$ in the case of $\Delta m^2_{23}< 0$]{The values
$m_{1,2,3}$ as functions of $\delta$ in the case of $\Delta
m^2_{23}< 0$}\label{m1m2m3ada1}
\vspace*{0.5cm}

\includegraphics[width=6.0cm, height=4.0cm]{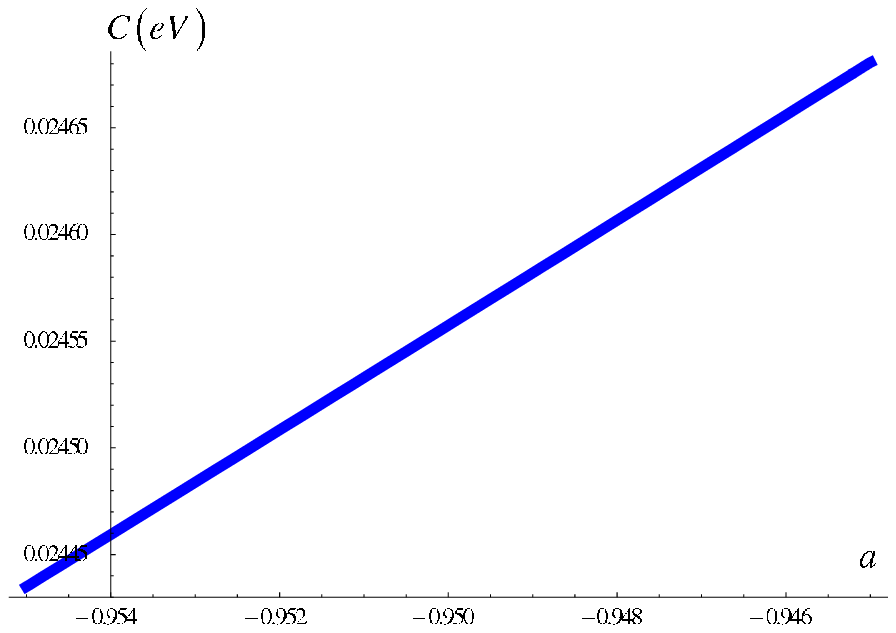}
\caption[The value $C$ as a function of  $a$ in the case of
$\Delta m^2_{23}< 0$]{The value $C$ as a function of  $a$ in the
case of $\Delta m^2_{23}< 0$}\label{Ca}
\end{center}
\end{figure}

In this case, the neutrino masses are functions of $\delta=\ep
v_\rho \tau$ as follows
 \bea
&&m_1=\frac{0.00364619}{\delta}+0.153928\,\delta,\label{m1pi}\\
&&m_2=\pm 8\times
10^{-3}\sqrt{18.7109+\frac{0.207729}{\delta^2}+370.216\, \delta^2},\label{m2pi}\\
&&m_3=\frac{0.00364619}{\delta}-0.153928\,\delta,\label{m3pi}\eea
In Fig. \ref{m123w1}, we have plotted the values $m_i \,
(i=1,2,3)$ as a function of $\delta$ with the values of $\delta\in
(-0.5, 0.5) \, \mathrm{eV}$. This figure shows that there exist
the allowed regions for value of $\delta$ where either inverted
($|m_1|\simeq |m_2|> |m_3|$) or quasi-degenerate neutrino masses
spectrum ($|m_1|\simeq |m_2| \simeq |m_3|$) achieved. The
quasi-degenerate mass hierarchy obtained when $\delta\rightarrow
0$ or $\delta\rightarrow \pm \infty$. The inverted mass hierarchy
is  obtained if $\delta$ takes the values around $(-0.20,-0.15)\,
\mathrm{eV}$ or (0.15, 0.20)\, $\mathrm{eV}$ as shown in Figs.
\ref{m123d1}a and \ref{m123d1}b, respectively. The Figs.
\ref{m123a1}a and \ref{m123a1}b give three absolute neutrino
masses $m_i$ with $\delta \in (-0.2, -0.15)$ and $\delta \in
(0.15, 0.20)$,
 respectively. The values $\sum^3_{i=1}m_i$ as well as $\sum^3_{i=1}|m_i|$
  as a functions of  $\delta$ are
  plotted in Figs. \ref{m1m2m3s1} and \ref{m1m2m3ada1}, respectively.
\\
\\
 In similarity to the normal case, to get explicit values of the
model parameters, we also assume $\delta \equiv \ep v_\rho \tau
=0.15\, \mathrm{eV}$,  which is safely small.
 Then the neutrino masses are explicitly
 given as \be
 m_1 \simeq 4.74\times 10^{-2}\, \mathrm{eV},\, m_2\simeq 4.82\times 10^{-2}\, \mathrm{eV},\,
  m_3\simeq1.22\times 10^{-3} \, \mathrm{eV},\label{numass1}\ee
 or \be
 m_1 \simeq 4.74\times 10^{-2}\, \mathrm{eV},\, m_2\simeq -4.82\times 10^{-2}\, \mathrm{eV},\,
  m_3\simeq1.22\times 10^{-3} \, \mathrm{eV}.\label{numass2}\ee
 From (\ref{numass1}) we find out
 \be
 A\simeq 4.76\times 10^{-2}\, \mathrm{eV},\, B\simeq 2.57\times 10^{-4} \,\mathrm{eV},\, C\simeq
 2.46\times 10^{-2}\, \mathrm{eV},\, D\simeq 2.34\times 10^{-2}\, \mathrm{eV}.\ee
  Furthermore, suppose that
\be \la_s =\la_{\si}=1\, \mathrm{eV},\, v_s=v_\si,\,
\La_s=\La_\si=v^2_s,\, \frac{v_\si}{\La_\si}=a\frac{v_s}{\La_s},
\label{vienrelat}\ee
 we obtain the relation between $C$ and $a$ as in Fig. \ref{Ca}. Then the satisfied value of
 $a$,  which can be inferred   from this figure,  is as follows  $a=-0.950$. With this value of $a$
  we get $x\simeq 1.22\times 10^{-2}$, $y \simeq 1.23\times 10^{-2}$, $z\simeq 1.11$.

In a similar way, from (\ref{numass2}) we get
$ A\simeq -1.85\times 10^{-2}\, \mathrm{eV},\, B\simeq -3.13\times 10^{-2} \,\mathrm{eV},\,
C\simeq 9.45\times 10^{-3}\, \mathrm{eV},\, D\simeq 8.23\times 10^{-3}\, \mathrm{eV}$.
 With the  assumption  in (\ref{vienrelat}) we get $a\simeq -0.9815$, and it follows $x\simeq -3.47
 \times 10^{-2},\, y \simeq 3.32\times 10^{-2}, \, z \simeq -9.12\times 10^{-3}$.

  In both case, the parameter $\tau \sim1.5\times 10^{-4}$ provided that $\ep \sim 10^{-3} \,
  \mathrm{eV}$ and $v_\rho \sim1.5\times 10^{-3} \mathrm{GeV}$.
  The solutions in (\ref{numass1}) and (\ref{numass2}) mean a inverted neutrino mass spectrum.

\section{\label{vev} Vacuum alignment}
In order to make this work completed we write out the scalar
potentials of the model.
 It is to be noted that $(\Tr{A})(\Tr{B})=\Tr{(A\Tr{B}})$
and we have used the following notation:
$V(\textit{X}\rightarrow\textit{X'},\textit{Y}\rightarrow\textit{Y'},\cdots)
\equiv V(X,Y,\cdots)\!\!\!\mid_{X=X',Y=Y',\cdots}$. The general
potential invariant under all subgroups takes the form:
\be V_{\mathrm{total}}=V_{\mathrm{tri}}+V_{\mathrm{sext}}+
V_{\mathrm{tri-sext}},\label{vien5}\ee
 where $V_{\mathrm{tri}}$
comes from only contributions of $\mathrm{SU}(3)_L$ triplets given
as a sum of: \bea V(\chi)&=&\mu_{\chi}^2\chi^\+\chi
+\lambda^{\chi}({\chi}^\+\chi)^2,\label{Vchi}\\
V(\phi)&=&V(\chi\rightarrow\phi), \hs
V(\phi')=V(\chi\rightarrow\phi'),\label{vphiphip}\hs
V(\eta)=V(\chi\rightarrow\eta),\crn
V(\eta')&=&V(\chi\rightarrow\eta'),\hs V(\rho)=V(\chi\rightarrow \rho),\label{vetaetaprho}\\
V(\phi,\chi)&=&\lambda_1^{\phi\chi}(\phi^\+\phi)(\chi^\+\chi)
+\lambda_2^{\phi\chi}(\phi^\+\chi)(\chi^\+\phi),\label{phichiint}\\
V(\phi,\phi')&=&V(\phi,\chi\rightarrow\phi')
+\lambda_3^{\phi\phi'}(\phi^\+\phi')(\phi^\+\phi')
+\lambda_4^{\phi\phi'}(\phi'^\+\phi)(\phi'^\+\phi),\crn
V(\phi,\eta)&=&V(\phi,\chi\rightarrow\eta),\hs
V(\phi,\eta')=V(\phi,\chi\rightarrow\eta'),\crn
V(\phi,\rho)&=&V(\phi,\chi\rightarrow \rho),\label{phirhoiint}\hs
V(\phi',\chi)=V(\phi\rightarrow\phi',\chi),\crn
V(\phi',\eta)&=&V(\phi\rightarrow\phi',\chi\rightarrow\eta),\hs
V(\phi',\eta')=V(\phi\rightarrow\phi',\chi\rightarrow\eta'),\crn
V(\phi',\rho)&=&V(\phi\rightarrow\phi',\chi\rightarrow\rho),\hs
V(\chi,\eta)=V(\phi\rightarrow\chi,\chi\rightarrow\eta),\crn
V(\chi,\eta')&=&V(\phi\rightarrow\chi,\chi\rightarrow\eta'),\hs
V(\chi,\rho) =V(\phi\rightarrow\chi,\chi\rightarrow\rho),\crn
V(\eta,\eta')&=&V(\phi\rightarrow\eta,\chi\rightarrow\eta')
+\lambda_3^{\eta\eta'}(\eta^\+\eta')(\eta^\+\eta')+\lambda_4^{\eta\eta'}(\eta'^\+\eta)(\eta'^\+\eta),\crn
V(\eta,\rho)&=&V(\phi\rightarrow\eta,\chi\rightarrow\rho),\hs
V(\eta',\rho)=V(\phi\rightarrow\eta',\chi\rightarrow\rho),\crn
V_{\chi\phi\phi'\eta\eta'\rho}&=&\mu_1\chi\phi\eta+\mu'_1\chi\phi'\eta'
+\la^1_1(\phi^\+\phi')(\eta^\+
\eta')+\la^2_1(\phi^\+\phi')(\eta'^\+ \eta)\crn
&+&\la^3_1(\phi^\+\eta)(\eta'^\+
\phi')+\la^4_1(\phi^\+\eta')(\eta^\+\phi')
+h.c.\label{vtrifourintract} \eea The $V_{\mathrm{sext}}$ is
summed over only antisextet contributions: \bea
V(s)&=&{\mu}^2_{s}\Tr(s^\+s)
+{\lambda}_1^{s}\Tr[(s^\+s)_{\underline{1}}(s^\+s)_{\underline{1}}]+
{\lambda}_2^{s}\Tr[(s^\+s)_{\underline{1'}}(s^\+s)_{\underline{1'}}]\crn
&+&{\lambda}_3^{s}\Tr[(s^\+s)_{\underline{1}''}(s^\+s)_{\underline{1}''}]
+{\lambda}_4^{s}\Tr[(s^\+s)_{\underline{1}'''}(s^\+s)_{\underline{1}'''}]
+{\lambda}_5^{s}\Tr(s^\+s)_{\underline{1}}\Tr(s^\+s)_{\underline{1}}\crn
&+&{\lambda}_6^{s}\Tr(s^\+s)_{\underline{1'}}\Tr(s^\+s)_{\underline{1'}}
+{\lambda}_7^{s}\Tr(s^\+s)_{\underline{1}''}\Tr(s^\+s)_{\underline{1}''}
+{\lambda}_8^{s}\Tr(s^\+s)_{\underline{1}'''}\Tr(s^\+s)_{\underline{1}'''},\label{vs}\\
V(\si)&=&{\mu}^2_{\si}\Tr(\si^\+\si)
+{\lambda}_1^{\si}\Tr[(\si^\+\si)_{\underline{1}}(\si^\+\si)_{\underline{1}}]
+{\lambda}_2^{\si}\Tr(\si^\+\si)_{\underline{1}}\Tr(\si^\+\si)_{\underline{1}},\label{vsi}\\
V(s,\si)&=&{\lambda}_1^{s\si}\Tr[(s^\+s)_{\underline{1}}(\si^\+\si)_{\underline{1}}]
+{\lambda}_2^{s\si}\Tr[(s^\+s)_{\underline{1}}]\Tr[(\si^\+\si)_{\underline{1}}]
\crn
&+&{\lambda}_3^{s\si}\Tr[(s^\+\si)_{\underline{2}}(\si^\+s)_{\underline{2}}]
+{\lambda}_{4}^{s\si}\Tr[(s^\+\si)_{\underline{2}}]\Tr[(\si^\+s)_{\underline{2}}]
+h.c.\nn \eea
The $V_{\mathrm{tri-sext}}$ is given as a sum over
all the terms connecting both the sectors:
\bea
V(\phi,s)&=&\lambda_1^{\phi
s}(\phi^\+\phi)\Tr(s^\+s)_{\underline{1}} +\lambda_2^{\phi
s}[(\phi^\+s^\+)(s\phi)]_{\underline{1}},\crn
V(\phi',s)&=&V(\phi\rightarrow\phi',s),\hs
V(\chi,s)=V(\phi\rightarrow\chi,s),\hs
V(\eta,s)=V(\phi\rightarrow\eta,s),\crn
V(\eta',s)&=&V(\phi\rightarrow\eta',s),\hs
V(\rho,s)=V(\phi\rightarrow\rho,s)+\{\la^{\rho s}_3\rho[(\rho
s^\+)s^\+]_{\underline{1'}}+h.c\},\crn
V(\phi,\si)&=&V(\phi,s\rightarrow \si)+\lambda_3^{\phi
\si}(\phi^\+\si)(\si^\+\phi)_{\underline{1}},\hs
V(\phi',\si)=V(\phi\rightarrow\phi', \si),\crn
V(\chi,\si)&=&V(\phi\rightarrow\chi,\si),\hs
V(\eta,\si)=V(\phi\rightarrow\eta,\si),\hs
V(\eta',\si)=V(\phi\rightarrow\eta',\si),\crn
V(\rho,\si)&=&V(\phi\rightarrow\rho,\si)+\{\la^{\rho
\si}_3\rho[(\rho \si^\+)\si^\+]_{\underline{1}'''}+h.c\},\crn
V(\phi,s,\si)&=&0,\hs V(\phi',s,\si)=0,\hs
V(\chi,s,\si)=0,\label{vanish1}\\
V(\eta,s,\si)&=&0,\hs
V(\eta',s,\si)=0,\hs V(\rho,s,\si)=0,\label{vanish2}\\
V_{s\si\chi\phi\phi'\eta\eta'\rho}&=&
(\lambda_1\phi^\+\phi'+\lambda_2\eta^\+\eta')\Tr(s^\+s)_{\underline{1'}}
+\lambda_3[(\phi^\+s^\+)(s\phi')]_{\underline{1}}
+\lambda_4[(\eta^\+s^\+)(s\eta')]_{\underline{1}}+h.c. \nn\eea
To provide the Majorana masses for the neutrinos, the lepton
number must be broken. This can be achieved via the scalar
potential violating $U(1)_{\mathcal{L}}$. However, the other
symmetries should be conserved. The violating $\mathcal{L}$
potential is given by \bea \bar{V}&=& [\overline{\lambda}_1\phi^\+
\phi+\overline{\lambda}_2\phi'^\+\phi'+\overline{\lambda}_3\chi^\+
\chi+\overline{\lambda}_4\eta^\+\eta
+\overline{\lambda}_5\eta'^\+\eta'+\overline{\lambda}_6\rho^\+\rho+\overline{\lambda}_7\eta^\+
\chi+\crn&& +\overline{\lambda}_8\Tr(s^\+s)_{\underline{1}}
+\overline{\lambda}_{9}\Tr(\si^\+\si)_{\underline{1}}](\eta^\+\chi)
+[\overline{\lambda}_{10}\phi^\+
\phi'+\overline{\lambda}_{11}\phi'^\+ \phi\crn&&
+\overline{\lambda}_{12}\eta^\+\eta'
+\overline{\lambda}_{13}\eta'^\+\eta
+\overline{\lambda}_{14}\eta'^\+\chi
+\overline{\lambda}_{15}\Tr(s^\+s)_{\underline{1'}}](\eta'^\+\chi)
+[\overline{\lambda}_{16}\eta^\+\phi
+\overline{\lambda}_{17}\eta'^\+\phi'
+\overline{\lambda}_{18}\eta'^\+\rho](\phi^\+\chi)\crn
&&+[\overline{\lambda}_{19} \eta^\+\phi'
+\overline{\lambda}_{20}\eta'^\+\phi](\phi'^\+\chi)
+\overline{\lambda}_{21}\Tr(s^\+s)_{\underline{1}'''}(\phi^\+\rho)
+\overline{\lambda}_{22}\Tr(s^\+s)_{\underline{1}''}(\phi'^\+\rho)
+\overline{\lambda}_{23}(\eta^\+s^\+)(s\chi)_{\underline{1}}\crn&&
+\overline{\lambda}_{24}[(\eta^\+\si^\+)(\si\chi)]_{\underline{1}}
+\overline{\lambda}_{25}[(\eta'^\+s^\+)(s\chi)]_{\underline{1}}
+\overline{\lambda}_{26}[(\rho^\+s^\+)(s\phi)]_{\underline{1}}
+\overline{\lambda}_{27}\phi[(\phi s^\+)s^\+]_{\underline{1}}
+\overline{\lambda}_{28}\phi[(\phi
\si^\+)\si^\+]_{\underline{1}}\crn&&
+\overline{\lambda}_{29}\phi'[(\phi' s^\+)s^\+]_{\underline{1'}}
+\overline{\lambda}_{30}\phi'[(\phi'
\si^\+)\si^\+]_{\underline{1'}}
+\overline{\lambda}_{31}\phi[(\phi' s^\+)s^\+]_{\underline{1}}
+\overline{\lambda}_{32}\phi'[(\phi
s^\+)s^\+]_{\underline{1'}}+h.c. \label{vbar} \eea In the
decomposing of $\underline{2}\otimes \underline{2}$,
$\underline{2}\otimes \underline{2}=\underline{1}\oplus
\underline{1}' \oplus \underline{1}{''}\oplus \underline{1}{'''}$,
there is no term which, as shown in (\ref{vanish1}),
(\ref{vanish2}), and (\ref{vbar}), is  invariant under combination
of  one scalar triplet and two different antisixtets; and some
couplings between $\rho$ and some other triplets are ruled out. As
a consequence, the general scalar potential violating
${\mathcal{L}}$ and  being  invariant   under $D_4$, is more
simpler than those of $S_3$ and $S_4$.

Let us now consider the potential $V_{tri}$. The flavons $\chi,
\phi, \phi', \eta,\eta'$ with their VEVs aligned in the same
direction (all of them are singlets) are an automatical solution
from the minimization conditions of $V_{tri}$. To explicitly see
this, in the system of equations for minimization, let us put
$v^*=v, v'^*=v', u^*=u, u'^*=u', v^*_\chi=v_\chi,
v^*_\rho=v_\rho$. Then  the potential minimization conditions for
triplets reduces to \bea \frac{\partial V_{tri}}{\partial
\om}&=&4\la^\chi\om^3+2\left(\mu^2_\chi+\la^{\chi\eta}_1
u^2+\la^{\chi\eta'}_1u'^2
+\la^{\chi\phi}_1v^2+\la^{\chi\phi'}_1v'^2+\la^{\chi\rho}_1v^2_\rho\right)
\om-\mu_1uv-\mu'_1u'v',\label{dhV1om}\\
 \frac{\partial V_{tri}}{\partial
v_\rho}&=&4\la^\rho v^3_\rho+2\left[\mu^2_\rho+\la^{\rho\eta}_1 u^2+\la^{\rho\eta'}_1u'^2
+(\la^{\rho\phi}_1+\la^{\rho\phi}_2)v^2+(\la^{\rho\phi'}_1+\la^{\rho\phi'}_2)v'^2
+\la^{\chi\rho}_1\om^2\right]v_\rho,\label{dhV1chi}\\
 \frac{\partial V_{tri}}{\partial v}&=&4\la^\phi v^3+2\left[\mu^2_\phi
 +\la^{\phi\eta}_1u^2+\la^{\phi\eta'}_1u'^2
+(\la^{\phi\phi'}_1+\la^{\phi\phi'}_2+\la^{\phi\phi'}_3+\la^{\phi\phi'}_4)v'^2
+(\la^{\rho\phi}_1+\la^{\rho\phi}_2)v^2_\rho
\right.\crn
&+&\left.\om^2\la^{\phi\chi}_1\right]v
+ (\la^1_1+\la^2_1)uu'v'-\mu_1\om u,\label{dhV1v}\\
   \frac{\partial V_{tri}}{\partial v'}&=&4\la^{\phi'} v'^3+2\left[\mu^2_{\phi'}
   +\la^{\phi'\eta}_1u^2+\la^{\phi'\eta'}_1u'^2
+(\la^{\phi\phi'}_1+\la^{\phi\phi'}_2+\la^{\phi\phi'}_3+\la^{\phi\phi'}_4)v^2
+(\la^{\rho\phi'}_1+\la^{\rho\phi'}_2)v^2_\rho
\right.\crn
&+&\left.\om^2\la^{\phi'\chi}_1\right]v'
+ (\la^1_1+\la^2_1)uu'v-\mu'_1\om u',\label{dhV1vp}\\
 \frac{\partial V_{tri}}{\partial u}&=&4\la^\eta u^3+2\left[\mu^2_\eta+(\la^{\eta\eta'}_1+
 \la^{\eta\eta'}_2+\la^{\eta\eta'}_3+\la^{\eta\eta'}_4)u'^2+\la^{\phi'\eta}_1v'^2
 +\la^{\phi\eta}_1v^2+\la^{\eta\rho}_1v^2_\rho+\om^2\la^{\eta\chi}_1\right]u
\crn
&+& (\la^1_1+\la^2_1)u'vv'-\mu_1\om v,\label{dhV1u}\\
 \frac{\partial V_{tri}}{\partial u'}&=&4\la^{\eta'} u'^3+2\left[\mu^2_{\eta'}+(\la^{\eta\eta'}_1
 +\la^{\eta\eta'}_2+\la^{\eta\eta'}_3+\la^{\eta\eta'}_4)u^2+\la^{\phi\eta'}_1v^2
 +\la^{\phi'\eta'}_1v'^2 +\la^{\eta'\rho}_1v^2_\rho+\om^2\la^{\eta'\chi}_1\right]u'\crn
 &+& (\la^1_1+\la^2_1)uvv'-\mu'_1\om v'.\label{dhV1up}
\eea It is easily to see that the derivatives of
$V_{tri}$ with respect to the variable $\om$ and
 $v_\rho$ showed in (\ref{dhV1om}), (\ref{dhV1chi}) are symmetric to each other. Similarly,
 the two pairs ($v, v'$) and ($u,u'$) behave the same as shown in Eqs.(\ref{dhV1v}),
 (\ref{dhV1vp}), (\ref{dhV1u}) and (\ref{dhV1up}). The parameters $\la^{\chi\phi}_{2}, \la^{\chi\phi'}_{2}$
 in Eq.(\ref{dhV1om}) vanish because of the interaction $(\phi^+\chi)(\chi^+\phi)$ in
 (\ref{phichiint}), and the parameters $\la^{\chi\phi}_{2,3,4}, \la^{\chi\phi'}_{2,3,4}$ in
 Eq.(\ref{dhV1om}), (\ref{dhV1chi}) vanish due to the symmetries of the model (such as $L$
 or $X$ or $\mathcal{L}$ or $D_4$ or one of their combinations).

The system of equations (\ref{dhV1om}) - (\ref{dhV1up}) always has
the solution ($u, v, u',v'$) as expected, even though the
complication. It is also noted that the above alignment is only
one of the solutions to be imposed to have the desirable results.
We have evaluated that the Eqs. (\ref{dhV1v}) - (\ref{dhV1up})
have the same structure solution. The solution is as follows
\bea
u=u'=v'=v=\pm\sqrt{\al}\label{special}\eea with \bea
\al&=&\left\{-\om^2(\la^{\chi\eta}_1+\la^{\chi\eta'}_1+\la^{\chi\phi}_1
+\la^{\chi\phi'}_1)+\om(\mu_1+\mu'_1)
-\mu^2_\eta-\mu^2_{\eta'}-\mu^2_\phi-\mu^2_{\phi'}\right.\crn
&+&\left.(\la^{\eta\rho}_1+\la^{\eta'\rho}_1+\la^{\phi\rho}_1+\la^{\phi'\rho}_1
+\la^{\phi\rho}_2+\la^{\phi'\rho}_2)v^2_\rho\right\}/\left\{2[\la^1_1+\la^2_1
+\la^{\eta\eta'}_1+\la^{\eta\eta'}_2+\la^{\eta\eta'}_3+\la^{\eta\eta'}_4\right.\crn
&+&\left.\la^{\phi\phi'}_1+\la^{\phi\phi'}_2+\la^{\phi\phi'}_3+\la^{\phi\phi'}_4
+\la^{\phi\eta}_1+\la^{\phi\eta'}_1+\la^{\phi'\eta}_1+\la^{\phi'\eta'}_1+\la^{\eta}+\la^{\eta'}
+\la^{\phi}+\la^{\phi'}]\right\}\crn
&\simeq&\frac{-2\om^2\la^{\chi\phi}_1+\om \mu_1 -2\mu^2_\phi
+3\la^{\phi\rho}_1v^2_\rho}
{2\la^1_1+12\la^{\phi\phi'}_1+4\la^{\phi}}.\label{vequal} \eea
Substituting  (\ref{vequal}) into (\ref{dhV1om}) and
(\ref{dhV1chi}) we obtain \bea \frac{\partial V_{tri}}{\partial
\om}&=&4\la^\chi\om^3+2\left(\mu^2_\chi+4\la^{\chi\phi}_1v^2+\la^{\chi\rho}_1v^2_\rho\right)\om
-2\mu_1v^2,\label{dhV1om1}\\
 \frac{\partial V_{tri}}{\partial
v_\rho}&=&4\la^\rho
v^3_\rho+2\left(\mu^2_\rho+6\la^{\rho\phi}_1v^2
+\la^{\chi\rho}_1\om^2\right)v_\rho,\label{dhV1chi1}\eea
  Noting that   the solution (\ref{special})  leads to special
relations among coupling constants: $\lambda^\eta = \lambda^{\eta'} = \lambda^\phi
= \lambda^{\phi'}  $ and mass parameters $\mu^2_{\eta} = \mu^2_{\eta'} =
\mu^2_{\phi} = \mu^2_{\phi'}$ and so on.  In general, these couplings and mass parameters
are independent, however,   the neutrino data and the discrete $D_4$ symmetry force them
being related.  This is the common  property of the discrete flavor symmetries.

Considering the potential $V_{sex}$ and $V_{tri-sex}$, we
urge that the contribution of $V_{\chi\phi\phi'\eta \eta'\rho}$ in
(\ref{vtrifourintract}) is very small in comparison with the other
terms in $V_{tri}$, so it can be neglected. From
(\ref{Vchi}) to (\ref{vtrifourintract}) and with the help of
(\ref{vevphiphip}), (\ref{vevchietaetap}), (\ref{rho}), and
imposing that \bea \la^*_1&=&\la_1,\, \la^*_2=\la_2, v^*_1=v_1,
v^*_2=v_2, \La^*_1=\La_1, \La^*_2=\La_2,\crn
\la^*_\si&=&\la_\si,\, v^*_\si=v_\si,\,\La^*_\si=\La_\si,\crn
v^*&=&v, v'^*=v', u^*=u, u'^*=u', v^*_\chi=v_\chi,
v^*_\rho=v_\rho, \nn \eea we obtain a system of equations of the potential
 minimization for anti-sextets:
\bea
\frac{\partial V_1}{\partial \la^*_1}&=&v^2_\chi\la^{\chi
s}_1\la_1+2(\la^s_{1}+\la^s_{2}+\la^s_{5}
+\la^s_{6})\la^3_1+4\la^s_{7}\la_{2}\La_{1}\La_{2}
+\la^{s\si}_{4}\La_{1}\La_{\si}\la_{\si}+2(\la^s_{1}+\la^s_{2})\La_{1}v^2_1\crn
&+&2(\la^s_{1}\la_{2}+\la^s_{1}\La_{2}-\la^s_{2}\la_{2}-\la^s_{2}\La_{2}
+3\la^s_{3}\la_{2}+\la^s_{3}\La_{2}+\la^s_{4}\la_{2}-\la^s_{4}\La_{2}+4\la^s_{7}\la_{2})v_1v_2\crn
&+&2(\la^s_3+\la^s_4)\La_{1}v^2_2+\left[(\la^{s\si}_{1}+\la^{s\si}_{3}+2\la^{s\si}_{4})\la_{\si}+
(\La^{s\si}_{1}+\la^{s\si}_{3})\La_{\si}\right]v_1v_\si\crn
&+&\la_{1}\left\{2(\la^s_5+\la^s_6)\La^2_1+(\la^{s\si}_1+\la^{s\si}_2+\la^{s\si}_3+\la^{s\si}_4)\la^2_\si
+\la^{s\si}_2\La^2_\si+\mu^2_s+(\la^{\eta s}_1+\la^{\eta
s}_2)u^2\right.\crn &+&\left.(\la^{\eta' s}_1+\la^{\eta'
s}_2)u'^2+\la^{\phi s}_1v^2+\la^{\phi' s}_1v'^2
+2\left[(\la^s_5-\la^s_6)\La^2_2+\la^2_2(\la^s_1-\la^s_2+2\la^s_3+\la^s_5\right.\right.\crn
&-&\left.\left.\la^s_6+2\la^s_7)+2(\la^s_1+\la^s_2+\la^s_5+\la^s_6)v^2_1
+(\la^s_1-\la^s_2+\la^s_3-\la^s_4+2\la^s_5-2\la^s_6)v^2_2\right]\right.\crn
&+&\left. \la^{\rho s}_1v^2_\rho+(\la^{\si s}_1+2\la^{\si
s}_2+\la^{\si s}_3)v^2_\si \right\},\label{parla1}\eea \bea
 \frac{\partial V_{1}}{\partial \la^*_2}&=&v^2_\chi\la^{\chi s}_1\la_2+2(\la^s_{1}+\la^s_{2}+\la^s_{5}
 +\la^s_{6})\la^3_2+4\la^s_{7}\la_{1}\La_{1}\La_{2}
+\la^{s\si}_{4}\La_{2}\La_{\si}\la_{\si}+2(\la^s_{1}+\la^s_{2})\La_{2}v^2_2\crn
&+&2(\la^s_{1}\la_{1}+\la^s_{1}\La_{1}-\la^s_{2}\la_{1}-\la^s_{2}\La_{1}
+3\la^s_{3}\la_{1}+\la^s_{3}\La_{1}+\la^s_{4}\la_{1}-\la^s_{4}\La_{1}+4\la^s_{7}\la_{1})v_1v_2\crn
&+&2(\la^s_3+\la^s_4)\La_{2}v^2_1+\left[(\la^{s\si}_{1}+\la^{s\si}_{3}+2\la^{s\si}_{4})\la_{\si}+
(\La^{s\si}_{1}+\la^{s\si}_{3})\La_{\si}\right]v_2v_\si\crn
&+&\la_{2}\left\{2(\la^s_5+\la^s_6)\La^2_2+(\la^{s\si}_1+\la^{s\si}_2+\la^{s\si}_3+\la^{s\si}_4)\la^2_\si
+\la^{s\si}_2\La^2_\si+\mu^2_s+(\la^{\eta s}_1+\la^{\eta s}_2)u^2\right.\crn
&+&\left.(\la^{\eta' s}_1+\la^{\eta' s}_2)u'^2+\la^{\phi s}_1v^2+\la^{\phi' s}_1v'^2
+2\left[(\la^s_5-\la^s_6)\La^2_1+\la^2_1(\la^s_1-\la^s_2+2\la^s_3+\la^s_5\right.\right.\crn
&-&\left.\left.\la^s_6+2\la^s_7)+2(\la^s_1+\la^s_2+\la^s_5+\la^s_6)v^2_2
+(\la^s_1-\la^s_2+\la^s_3-\la^s_4+2\la^s_5-2\la^s_6)v^2_1\right]\right.\crn
&+&\left. \la^{\rho s}_1v^2_\rho+(\la^{\si s}_1+2\la^{\si s}_2+\la^{\si s}_3)v^2_\si \right\},\label{parla2}\eea
\bea
 \frac{\partial V_{1}}{\partial v^*_1}&=&\left[v^2_\chi(2\la^{\chi s}_1+\la^{\chi s}_2)+2\mu^2_s+(2\la^{\eta s}_1
 +\la^{\eta s}_2)u^2+(2\la^{\eta' s}_1+\la^{\eta' s}_2)u'^2+2\la^{\phi s}_1v^2+2\la^{\phi' s}_1v'^2\right]v_1\crn
&-&2\la^s_4(\la_2-\La_2)\left[(\la_2-\La_2)v_1-(\la_1-\La_1)v_2\right]+
8\la^s_7v_2(\la_1\la_2+\La_1\La_2+2v_1v_2)\crn
&+&4\la^s_6v_1\left[(\la^2_1-\la^2_2)+(\La^2_1-\La^2_2)+2(v^2_1-v^2_2)\right]
+2\left\{v_1\left[(\la^2_2+\La^2_2)(\la^s_1-\la^s_2)\right.\right.\crn
&+&\left.\left.2(\la^2_1+\la_1\La_1+\La^2_1+v^2_1)(\la^s_1+\la^s_2)\right]
+(\la_1+\La_1)(\la_2+\La_2)(\la^s_1-\la^s_2)v_2+2(\la^s_1-\la^s_2)v_1v^2_2\right\}\crn
&+&2\la_3\left\{(\la_2+\La_2)^2v_1+\left[\la_1(3\la_2+\La_2)+\La_1(\la_2
+3\La_2)\right]v_2+4v_1v^2_2\right\}\crn
&+&4\la^s_5v_1\left[\la^2_1+\la^2_2+\La^2_1+\La^2_2+2(v^2_1+v^2_2)\right]+2\la^{\rho s}_1v_1v^2_\rho
+2\la^{s\si}_4v_\si (\la_1\la_\si+\La_1\La_\si+2v_1v_\si)\crn
&+&2\la^{s\si}_2v_1 (\la^2_\si+\La^2_\si+2v^2_\si)
+(\la^{s\si}_1+\la^{s\si}_3) \left[(\la^2_\si+\La^2_\si)v_1+(\la_1
+\La_1)(\la_\si+\La_\si)v_\si+2v_1v^2_\si\right],\crn
\label{parv1}\eea
\bea
 \frac{\partial V_{1}}{\partial v^*_2}&=&\left[v^2_\chi(2\la^{\chi s}_1+\la^{\chi s}_2)
 +2\mu^2_s+(2\la^{\eta s}_1+\la^{\eta s}_2)u^2+(2\la^{\eta' s}_1+\la^{\eta' s}_2)u'^2
 +2\la^{\phi s}_1v^2+2\la^{\phi' s}_1v'^2\right]v_2\crn
&-&2\la^s_4(\la_1-\La_1)\left[(\la_1-\La_1)v_2-(\la_2-\La_2)v_1\right]
+8\la^s_7v_1(\la_1\la_2+\La_1\La_2+2v_1v_2)\crn
&+&4\la^s_6v_2\left[(\la^2_2-\la^2_1)+(\La^2_2-\La^2_1)+2(v^2_2-v^2_1)\right]
+2\left\{v_2\left[(\la^2_1+\La^2_1)(\la^s_1-\la^s_2)\right.\right.\crn
&+&\left.\left.2(\la^2_2+\la_2\La_2+\La^2_2+v^2_2)(\la^s_1+\la^s_2)\right]
+(\la_1+\La_1)(\la_2+\La_2)(\la^s_1-\la^s_2)v_1+2(\la^s_1-\la^s_2)v_2v^2_1\right\}\crn
&+&2\la_3\left\{(\la_1+\La_1)^2v_2+\left[\la_1(3\la_2+\La_2)
+\La_1(\la_2+3\La_2)\right]v_1+4v_2v^2_1\right\}\crn
&+&4\la^s_5v_2\left[\la^2_1+\la^2_2+\La^2_1+\La^2_2
+2(v^2_1+v^2_2)\right]+2\la^{\rho s}_1v_2v^2_\rho+2\la^{s
\si}_4v_\si (\la_2\la_\si+\La_2\La_\si+2v_2v_\si)\crn
&+&2\la^{s\si}_2v_2 (\la^2_\si+\La^2_\si+2v^2_\si)
+(\la^{s\si}_1+\la^{s\si}_3) \left[(\la^2_\si+\La^2_\si)v_2
+(\la_2+\La_2)(\la_\si+\La_\si)v_\si+2v_2v^2_\si\right],\crn
\label{parv2}\eea
 \bea \frac{\partial V_{1}}{\partial \La^*_1}&=&v^2_\chi(\la^{\chi s}_1
 +\la^{\chi s}_2)\La_1+2(\la^s_{1}+\la^s_{2}
 +\la^s_{5}+\la^s_{6})\La^3_1+4\la^s_{7}\la_{2}\la_{1}\La_{2}
+\la^{s\si}_{4}\la_{1}\La_{\si}\la_{\si}+2(\la^s_{1}+\la^s_{2})\la_{1}v^2_1\crn
&+&2(\la^s_{1}\la_{2}+\la^s_{1}\La_{2}-\la^s_{2}\la_{2}-\la^s_{2}\La_{2}
+3\la^s_{3}\La_{2}+\la^s_{3}\la_{2}+\la^s_{4}\La_{2}-\la^s_{4}\la_{2}+4\la^s_{7}\La_{2})v_1v_2\crn
&+&2(\la^s_3+\la^s_4)\la_{1}v^2_2+\left[(\la^{s\si}_{1}+\la^{s\si}_{3}+2\la^{s\si}_{4})\La_{\si}+
(\La^{s\si}_{1}+\la^{s\si}_{3})\la_{\si}\right]v_1v_\si\crn
&+&\La_{1}\left\{2(\la^s_5-\la^s_6)\la^2_2
+(\la^{s\si}_1+\la^{s\si}_3+\la^{s\si}_4)\La^2_\si
+\mu^2_s+\la^{\eta s}_1u^2+\la^{\eta' s}_1u'^2+\la^{\phi
s}_1v^2+\la^{\phi' s}_1v'^2\right.\crn &+&\left.
2\left[\La^2_2(\la^s_1-\la^s_2+2\la^s_3+\la^s_5-\la^s_6
+2\la^s_7)+2(\la^s_1+\la^s_2+\la^s_5+\la^s_6)v^2_1
+(\la^s_1-\la^s_2+\la^s_3 \right.\right.\crn
&-&\left.\left.\la^s_4+2\la^s_5-2\la^s_6)v^2_2\right]+ \la^{\rho
s}_1v^2_\rho +(\la^{\si s}_1+\la^{\si s}_3)v^2_\si+\la^{\si
s}_2(\la^2_\si+\La^2_\si+2v^2_\si) \right\},\label{parLa1}\eea
 \bea \frac{\partial V_{1}}{\partial \La^*_2}&=&v^2_\chi(\la^{\chi s}_1
 +\la^{\chi s}_2)\La_2+2(\la^s_{1}+\la^s_{2}
 +\la^s_{5}+\la^s_{6})\La^3_2+4\la^s_{7}\la_{2}\la_{1}\La_{1}
+\la^{s\si}_{4}\la_{2}\La_{\si}\la_{\si}+2(\la^s_{3}+\la^s_{4})\la_{2}v^2_1\crn
&+&2(\la^s_{1}\la_{1}+\la^s_{1}\La_{1}-\la^s_{2}\la_{1}-\la^s_{2}\La_{1}
+3\la^s_{3}\La_{1}+\la^s_{3}\la_{1}+\la^s_{4}\La_{1}-\la^s_{4}\la_{1}
+4\la^s_{7}\La_{1})v_1v_2\crn
&+&2(\la^s_1+\la^s_2)\la_{2}v^2_2+\left[(\la^{s\si}_{1}
+\la^{s\si}_{3}+2\la^{s\si}_{4})\La_{\si}+
(\La^{s\si}_{1}+\la^{s\si}_{3})\la_{\si}\right]v_2v_\si\crn
&+&\La_{2}\left\{2(\la^s_5-\la^s_6)\la^2_1
+(\la^{s\si}_1+\la^{s\si}_3+\la^{s\si}_4)\La^2_\si
+\mu^2_s+\la^{\eta s}_1u^2+\la^{\eta' s}_1u'^2+\la^{\phi
s}_1v^2+\la^{\phi' s}_1v'^2\right.\crn &+&\left.
2\left[\La^2_1(\la^s_1-\la^s_2+2\la^s_3+\la^s_5-\la^s_6
+2\la^s_7)+2(\la^s_1+\la^s_2+\la^s_5+\la^s_6)v^2_2
+(\la^s_1-\la^s_2+\la^s_3 \right.\right.\crn
&-&\left.\left.\la^s_4+2\la^s_5-2\la^s_6)v^2_1\right]+ \la^{\rho
s}_1v^2_\rho +(\la^{\si s}_1+\la^{\si s}_3)v^2_\si+\la^{\si
s}_2(\la^2_\si+\La^2_\si+2v^2_\si) \right\},\label{parLa2} \eea
where $V_{1}$ is a sum of $V_{sext}$ and $V_{tri-sext}$: \be
V_{1}=V_{sext}+V_{tri-sext} \ee It is easily to see that the
equations (\ref{parla1}), (\ref{parla2}), (\ref{parv1}),
(\ref{parv2}), (\ref{parLa1}) and (\ref{parLa2}) take the same
form in couples. This system of equations yields the following
solutions \be \la_1 = \beta \la_2,\hs v_1=\beta v_2,\hs
\La_1=\beta \La_2, \label{solutions} \ee where $\beta$ is a
constant. It means that there are several alignments for VEVs. In
this work, to have the desirable results, we have imposed the two
directions for breaking $D_4 \rightarrow Z_2 \otimes Z_2$ and
$D_4\rightarrow Z_2$ as mentioned, in which $\beta =1$ and
$\beta\neq1$ but is approximate to the unit. In the case that
$\beta =1$ or $\langle s_1\rangle=\langle s_2\rangle$, we have
\bea \frac{\partial V_1}{\partial \la_1}&=&\frac{\partial
V_1}{\partial \la_2}\equiv \frac{\partial V_1}{\partial \la_s},\,
\frac{\partial V_1}{\partial v_1}=\frac{\partial V_1}{\partial
v_2}\equiv \frac{\partial V_1}{\partial v_s},\, \frac{\partial
V_1}{\partial \La_1}=\frac{\partial V_1}{\partial \La_2}\equiv
\frac{\partial V_1}{\partial \La_s},\eea and this system reduces
to \bea \frac{\partial V_1}{\partial \la_s}
&=&4(\la^s_{1}+\la^s_{3}+\la^s_{5}+\la^s_{7})\la^3_s
+\la_{s}\left[\mu^2_s+4(\la^s_5+\la^s_7)\La^2_s+(\la^{s\si}_1
+\la^{s\si}_2+\la^{s\si}_3+\la^{s\si}_4)\la^2_\si
\right.\crn
&+&\left.\la^{s\si}_2\La^2_\si+8(\la^s_1+\la^s_3+\la^s_5+\la^s_7)v^2_s+(\la^{\si
s}_1+2\la^{\si s}_2+ \la^{\si s}_3)v^2_\si
\right]+4(\la^s_{1}+\la^s_{3})\La_{s}v^2_s\crn
&+&(\la^{s\si}_{1}+\la^{s\si}_{3})(\la_{\si}+\La_{\si})v_sv_\si+\la^{s\si}_{4}(\La_{s}\La_{\si}
+2v_sv_\si)\la_{\si}, \label{parlasequ}\\
 \frac{\partial V_1}{\partial v_s}
&=&4(\la^s_{1}+\la^s_{3}+2\la^s_{5}+2\la^s_{7})v^3_s
+v_{s}\left[2\mu^2_s+8(\la^s_1+\la^s_3+\la^s_5+\la^s_7)(\la^2_s+\La^2_s)\right.\crn
&+&\left.(\la^{s\si}_1+2\la^{s\si}_2+\la^{s\si}_3)(\la^2_\si +\La^2_\si)+8(\la^s_{1}+\la^s_{3})\la_s\La_s
+2(\la^{\si s}_1+2\la^{\si s}_2+\la^{\si s}_3+2\la^{\si s}_4)v^2_\si\right]\crn
&+&\left[(\la^{s\si}_{1}+\la^{s\si}_{3})(\la_{\si}+\La_{\si})(\la_{s}+\La_{s})
+2\la^{s\si}_4(\la_s\La_s+\la_\si\La_\si)\right]v_\si, \label{parvsequ}\\
 \frac{\partial V_1}{\partial \La_s}
&=&4(\la^s_{1}+\la^s_{3}+\la^s_{5}+\la^s_{7})\La^3_s
+\La_{s}\left[\mu^2_s+4(\la^s_5+\la^s_7)\la^2_s+(\la^{s\si}_1+\la^{s\si}_2
+\la^{s\si}_3+\la^{s\si}_4)\La^2_\si \right.\crn
&+&\la^{s\si}_2\la^2_\si+\left.8(\la^s_1+\la^s_3+\la^s_5+\la^s_7)v^2_s+(\la^{\si s}_1
+2\la^{\si s}_2+\la^{\si s}_3)v^2_\si\right]
+4(\la^s_{1}+\la^s_{3})\la_{s}v^2_s\crn
&+&(\la^{s\si}_{1}+\la^{s\si}_{3})(\la_{\si}+\La_{\si})v_sv_\si+\la^{s\si}_{4}(\la_{s}\la_{\si}
+2v_sv_\si)\La_{\si}. \label{parLasequ}\eea
The derivatives of $V_{1}$ with respect to the variable $\la_s$ and $\La_s$ as shown in (\ref{parlasequ}),
(\ref{parLasequ}) are symmetric to each other.

\section{Gauge bosons \label{Gaugeboson}}
The covariant derivative of a triplet is given by
\be
D_{\mu} = \partial_{\mu}-ig\frac{\lambda_a}{2}W_{\mu a}-ig_XX
\frac{\lambda_9}{2}B_{\mu}=\partial_{\mu}-iP_{\mu},\hs\label{331rhcova}
\ee
where $\lambda_a (a=1,2,...,8)$ are Gell-Mann matrices,
$\lambda_9=\sqrt{\frac{2}{3}}\mathrm{diag}(1,1,1)$,
 $\mathrm{Tr}{\lambda_a\lambda_b}=2\delta_{ab}$,
 $\mathrm{Tr}{\lambda_9\lambda_9}=2$, and $X$ is $X$ -charged of Higgs triplets.
Let us denote the following combinations:
\bea
W'^+_{\mu}&=&\frac{W_{\mu 1}-iW_{\mu 2}}{\sqrt{2}},\, X'^0_{\mu}=\frac{W_{\mu 4}-iW_{\mu 5}}{\sqrt{2}},\crn
Y'^-_{\mu}&=&\frac{W_{\mu 6}-iW_{\mu 7}}{\sqrt{2}}, \,
W'^-_{\mu}=(W'^{+}_{\mu})^*,\, Y'^+_{\mu}=(Y'^{-}_{\mu})^*,\label{WYXt}
\eea
then $P_{\mu}$ is rewritten in a convenient form as follows:
\be
\frac{g}{2}\left(%
\begin{array}{ccc}
W_{\mu 3}+\frac{W_{\mu 8}}{\sqrt{3}}+t\sqrt{\frac{2}{3}}XB_{\mu}&\sqrt{2}W'^+_{\mu}&\sqrt{2}X'^0_{\mu},\\
\sqrt{2}W'^-_{\mu}&-W_{\mu 3}+\frac{W_{\mu 8}}{\sqrt{3}}+t\sqrt{\frac{2}{3}}XB_{\mu}&\sqrt{2}Y'^-_{\mu}\\
\sqrt{2}X'^{0*}_{\mu}&\sqrt{2}Y'^+_{\mu}&-\frac{2}{\sqrt{3}}W_{\mu 8}+t\sqrt{\frac{2}{3}}XB_{\mu}\\
\end{array}%
\right), \label{pmu}
\ee
with \[ t = \fr{ g_X}{g}.\]
We note that $W_4$ and $W_5$ are pure real and imaginary parts of $X^0$ and $X^{0*}$, respectively.

The covariant derivative for an antisextet with the VEV part is \cite{DongLongdh, DongHLT}
\be
D_{\mu}\langle s_i\rangle =\frac{ig}{2}\{W_{\mu}^a\la^*_a\langle s_i\rangle+\langle s_i\rangle
W_{\mu}^a\la^{*T}_a\}+ig_X T_9 X B_\mu \langle s_i\rangle.\label{antisextcova}
\ee
The covariant derivative (\ref{antisextcova}) acting on the antisextet VEVs are given by
\bea
{[D_{\mu}\langle s_i\rangle]_{11}}&=&ig\left(\la_i W_{\mu 3}+\frac{\la_i}{\sqrt{3}}W_{\mu 8}
+\frac{1}{3}\sqrt{\frac{2}{3}}t\la_i B_\mu + \sqrt{2}v_i X'^{0*}\right),\crn
{[D_{\mu}\langle s_i\rangle]_{12}}&=&\frac{ig}{\sqrt{2}}\left(\la_i W'^+_{\mu}+v_i Y'^+_{\mu}\right),\crn
{[D_{\mu}\langle s_i\rangle]_{13}}&=&\frac{ig}{2}\left(v_i W_{\mu 3}-\frac{v_i}{\sqrt{3}} W_{\mu8}
+\frac{2}{3}\sqrt{\frac{2}{3}}tv_i B_\mu +\sqrt{2}\la_i X'^0_\mu+\sqrt{2}\La_i X'^{0*}_\mu\right),\crn
{[D_{\mu}\langle s_i\rangle]_{22}}&=&0,\hs
{[D_{\mu}\langle s_i\rangle]_{23}}=\frac{ig}{\sqrt{2}}\left(v_i W'^+_{\mu}+\La_i Y'^+_{\mu}\right),\crn
{[D_{\mu}\langle s_i\rangle]_{33}}&=&ig\left(-\frac{2}{\sqrt{3}}\La_i W_{\mu8}+\frac{1}{3}
\sqrt{\frac{2}{3}}t\La_i B_\mu +\sqrt{2}v_i X'^0_\mu\right),\crn
{[D_{\mu}\langle s_i\rangle]_{21}}&=&{[D_{\mu}\langle s_i\rangle]_{12}},\hs {[D_{\mu}
\langle s_i\rangle]_{31}}={[D_{\mu}\langle s_i\rangle]_{13}},\hs
{[D_{\mu}\langle s_i\rangle]_{32}}={[D_{\mu}\langle s_i\rangle]_{23}}.\nn
\eea
The masses of gauge bosons in this model are defined
\bea
\mathcal{L}^{GB}_{mass}&=&(D_{\mu}\langle\phi\rangle)^+(D^{\mu}\langle\phi\rangle)
+(D_{\mu}\langle\phi'\rangle)^+(D^{\mu}\langle\phi'\rangle)+(D_{\mu}\langle\chi\rangle)^+
(D^{\mu}\langle\chi\rangle) \crn
&+&(D_{\mu}\langle\eta\rangle)^+(D^{\mu}\langle\eta\rangle)
+(D_{\mu}\langle\eta'\rangle)^+(D^{\mu}\langle\eta'\rangle)
+(D_{\mu}\langle\rho\rangle)^+(D^{\mu}\langle\rho\rangle) \crn
&+&\Tr[(D_{\mu}\langle s_1\rangle)^+(D^{\mu}\langle s_1\rangle)]+\Tr[(D_{\mu}\langle
s_2\rangle)^+(D^{\mu}\langle s_2\rangle)]
+\Tr[(D_{\mu}\langle \si \rangle)^+(D^{\mu}\langle \si \rangle)],\crn
\label{LGBfull}\eea
where $\mathcal{L}^{GB}_{mass}$ in (\ref{LGBfull}) is different from  one in Ref. \cite{DongHLT}
 by the contribution from  the $\rho$ and the term relating to the anti-sextet $\si$. In Ref. \cite{DongHLT}
  the $\rho$ and $s'$ contributions were skip at the first order. In the following, we note that
  $\langle s_1\rangle =\langle s_1\rangle$, namely $\la_1 =\la_2=\la_s, v_1=v_2=v_s,
   \La_1 =\La_2=\La_s$ are taken
into account.

Substitute the Higgs VEVs of the model from (\ref{vevphiphip}), (\ref{vevchietaetap}),
(\ref{vevsi}),  (\ref{vevs})  and (\ref{vevrho}) into (\ref{LGBfull}) we obtain
\bea
\mathcal{L}^{GB}_{mass}&=&
\frac{v^2}{324}\left[81g^2(W^2_{\mu 1}+W^2_{\mu 2})+81g^2(W^2_{\mu 6}+W^2_{\mu 7})
+(-9gW_{\mu3}+3\sqrt{3}gW_{\mu8}+2\sqrt{6}g_X B_\mu)^2\right]\crn
&+&\frac{v'^2}{324}\left[81g^2(W^2_{\mu 1}+W^2_{\mu 2})+81g^2(W^2_{\mu 6}+W^2_{\mu 7})
+(-9gW_{\mu3}+3\sqrt{3}gW_{\mu8}+2\sqrt{6}g_X B_\mu)^2\right]\crn
&+&\frac{\om^2}{108}\left[27g^2(W^2_{\mu 4}+W^2_{\mu 5})+27g^2(W^2_{\mu 6}+W^2_{\mu7})
+36g^2W^2_{\mu 8}+12\sqrt{2}gg_xW_{\mu 8}B_\mu +2g^2_X B^2_\mu\right]\crn
&+&\frac{u^2}{324}\left[81g^2(W^2_{\mu 1}+W^2_{\mu 2})+81g^2(W^2_{\mu 4}+W^2_{\mu 5})
+(-9gW_{\mu3}-3\sqrt{3}gW_{\mu8}+\sqrt{6}g_X B_\mu)^2\right]\crn
&+&\frac{u'^2}{324}\left[81g^2(W^2_{\mu 1}+W^2_{\mu 2})+81g^2(W^2_{\mu 4}+W^2_{\mu 5})
+(-9gW_{\mu3}-3\sqrt{3}gW_{\mu8}+\sqrt{6}g_X B_\mu)^2\right]\crn
&+&\frac{v_\rho^2}{324}\left[81g^2(W^2_{\mu 1}+W^2_{\mu 2})+81g^2(W^2_{\mu 6}+W^2_{\mu7})
+(-9gW_{\mu3}+3\sqrt{3}gW_{\mu8}+2\sqrt{6}g_X B_\mu)^2\right]\crn
&+&
\frac{g^2}{6}\left[2(2\La_s v_s+\La_\si v_\si)\left(3W_{\mu 3}W_{\mu 4}+3W_{\mu 1}W_{\mu 6}-
3W_{\mu 2}W_{\mu 7}-5\sqrt{3}W_{\mu 4}W_{\mu 8}\right)\right.\crn
&+&\left.3(2v^2_s+v^2_\si+2\la^2_s+\la^2_\si)(W^2_{\mu1}+W^2_{\mu2})
+3(2v^2_s+v^2_\si+4\la^2_s+2\la^2_\si)W^2_{\mu3}\right.\crn
&+&\left.3(8v^2_s+4v^2_\si+2\la^2_s+\la^2_\si+2\La^2_s+\La^2_\si
+4\La_s\la_s+2\La_\si\la_\si)W^2_{\mu4}\right.\crn
&+&\left.3\left(8v^2_s+4v^2_\si+2\la^2_s+\la^2_\si+2\La^2_s
+\La^2_\si-4\La_s\la_s-2\La_\si\la_\si\right)W^2_{\mu5}\right.\crn
&+&\left.3(2v^2_s+v^2_\si+2\La^2_s+\La^2_\si)W^2_{\mu6}
+3(2v^2_s+v^2_\si+2\La^2_s+\La^2_\si)W^2_{\mu7}\right.\crn
&+&\left.2\sqrt{3}(-2v^2_s-v^2_\si+4\la^2_s+2\la^2_\si)W_{\mu3}W_{\mu8}
+(2v^2_s+v^2_\si+4\la^2_s+2\la^2_\si+18\La^2_s+8\La^2_\si)W^2_{\mu8}\right.\crn
&+&\left.18(2\la_s v_s+\la_\si v_\si)W_{\mu3}W_{\mu4}+6(2\la_s v_s+\la_\si v_\si)W_{\mu1}W_{\mu6}
-6(2\la_s v_s+\la_\si v_\si)W_{\mu2}W_{\mu7}\right.\crn
&+&\left. 2\sqrt{3}(2\la_s v_s+\la_\si v_\si)W_{\mu4}W_{\mu8}\right]\crn
&+&\frac{2}{27}t^2g^2(2\la^2_s+\la^2_\si+2\La^2_s+\La^2_\si+4v^2_s+2v^2_\si)B^2_{\mu}
-\frac{2\sqrt{6}}{9}tg^2(2\la^2_s+\la^2_\si+2v^2_s+v^2_\si)W_{\mu3}B_\mu\crn
&-&\frac{4\sqrt{6}}{9}tg^2\left[(2\la_s+2\La_s)v_s+(\la_\si+\La_\si)v_\si\right]W_{\mu4}B_\mu \crn
&+&\frac{2\sqrt{2}}{9}tg^2(2v^2_s+v^2_\si+4\La^2_s
+2\La^2_\si-2\la^2_s-\la^2_\si)W_{\mu8}B_\mu.\label{LGB}\eea
We can separate $\mathcal{L}^{GB}_{mass}$ in (\ref{LGB}) into
\be
\mathcal{L}^{GB}_{mass}=\mathcal{L}^{W_5}_{mass}
+\mathcal{L}^{CGB}_{mix}+\mathcal{L}^{NGB}_{mix},\label{LGBsplit}
\ee
where $\mathcal{L}^{W_5}_{mass}$ is the Lagrangian
of the imaginary part $W_{5}$. This boson is decoupled with mass given by
\be
M^2_{W_{5}}=
\frac{g^2}{2}\left(\om^2+u^2+u'^2+16v^2_s+8v^2_\si+4\la^2_s
+2\la^2_\si+4\La^2_s+2\La^2_\si-8\La_s\la_s
-4\La_\si\la_\si\right).\label{mW5}\ee
In the limit $\la_s, \la_\si, v_s, v_\si \rightarrow 0$, $M^2_{W_{5}}$ reduces to
\be
M^2_{W_{5}} =
\frac{g^2}{2}\left(\om^2+u^2+u'^2+4\La^2_s+2\La^2_\si\right).\label{mW5limit}\ee
$\mathcal{L}^{CGB}_{mix}$ is the Lagrangian part of the charged gauge bosons $W$ and $Y$,
\bea
\mathcal{L}^{CGB}_{mix}&=&
\frac{g^2}{4}(v^2+v'^2+\om^2+u^2+u'^2+v^2_\rho)\left(W^2_{\mu 1}
+W^2_{\mu 2}+W^2_{\mu 6}+W^2_{\mu 7}\right)\crn
&+&\frac{g^2}{6}\left[2(2\La_s v_s+\La_\si v_\si)\left(3W_{\mu 1}W_{\mu 6}
-3W_{\mu 2}W_{\mu 7}\right)\right.\crn
&+&\left.3(2v^2_s+v^2_\si+2\la^2_s+\la^2_\si)W^2_{\mu1}+3(2v^2_s+v^2_\si
+2\la^2_s+\la^2_\si)W^2_{\mu2}\right.\crn
&+&\left.3(2v^2_s+v^2_\si+2\La^2_s+\La^2_\si)W^2_{\mu6}+3(2v^2_s
+v^2_\si+2\La^2_s+\La^2_\si)W^2_{\mu7}\right.\crn
&+&\left.6(2\la_s v_s+\la_\si v_\si)W_{\mu1}W_{\mu6}-6(2\la_s v_s
+\la_\si v_\si)W_{\mu2}W_{\mu7}\right].\label{LCGB1}
\eea
We can rewrite $\mathcal{L}^{CGB}_{mix}$ in matrix form
\[
\mathcal{L}^{CGB}_{mix}
=\frac{g^2}{4}(W'^-_{\mu}\hs Y'^-_{\mu})M^2_{WY}\left( W'^{+ \mu} \hs
  Y'^{+ \mu} \right)^T,
\]
where
\be
M^2_{WY}=\left(%
\begin{array}{cc}
 m^2_{11}&m^2_{12} \\
 m^2_{21}&m^2_{22}\\
\end{array}%
\right),\label{MCGB}
\ee
with
\bea
m^2_{11}&=&2(v^2+v'^2
+u^2+u'^2+v_\rho^2+4v^2_s+2v^2_\si+4\la^2_s+2\la^2_\si),\crn
m^2_{12}&=&m^2_{21}=4(2\La_s v_s+2\la_s v_s+\La_\si v_\si+\la_\si v_\si),\crn
m^2_{22}&=&2(v^2+v'^2+\om^2+v_\rho^2+4v^2_s+2v^2_\si
+4\La^2_s+2\La^2_\si).\label{MCGBelement}
\eea
The matrix $M^2_{WY}$ in (\ref{MCGB}) with the elements in
(\ref{MCGBelement}) can be diagonalized as follows
\[
U^T_2M^2_{WY}U_2 = \mathrm{diag}(M^2_W, M^2_Y),
\]
where
\bea
M^2_W&=&\frac{g^2}{4}\left\{2v^2+2v'^2+u^2+u'^2+\om^2+2v^2_\rho+4\la^2_s+4\La^2_s+8v^2_s+
2\la^2_\si+2\La^2_\si+4v^2_\si-\sqrt{\Ga}\right\},\crn
M^2_Y&=&\frac{g^2}{4}\left\{2v^2+2v'^2+u^2+u'^2+\om^2+2v^2_\rho+4\la^2_s+4\La^2_s+8v^2_s+
2\la^2_\si+2\La^2_\si+4v^2_\si+\sqrt{\Ga}\right\},\crn
\label{MWY}\eea
with
\bea
\Ga&=&16\la^4_s+16\La^4_s+(2\la^2_\si-2\La^2_\si-\om^2+u^2+u'^2)^2\crn
&-&8\la^2_s(4\La^2_s-2\la^2_\si+2\La^2_\si+\om^2-u^2-u'^2-8v^2_s)\crn
&-&8\La^2_s(2\la^2_\si-2\La^2_\si-\om^2+u^2+u'^2-8v^2_s)
+64\La_s(\la_\si+\La_\si)v_s v_\si\crn
&+&16(\la_\si+\La_\si)^2v^2_\si+64\la_s v_s(2\La_s v_s
+\la_\si v_\si +\La_\si v_\si). \label{Ga}\eea
With corresponding eigenstates,  the charged gauge boson mixing matrix takes the form:
\[
U_2 = \left(%
\begin{array}{cc}
  \cos\theta&-\sin\theta \\
 \sin\theta&\cos\theta\\
\end{array}%
\right),
\]
where, the mixing angle $\theta$ is given by
\be
\tan\theta =\frac{8(\la_s+\La_s)v_s+4(\la_\si+\La_\si)v_\si}{4\la^2_s-
4\La^2_s+2\la^2_\si-2\La^2_\si-\om^2+u^2+u'^2-\sqrt{\Ga}}.
\label{theta}\ee
The physical charged gauge bosons is defined
\bea
W^-_\mu&=&\cos\theta W'^-_\mu+\sin\theta Y'^-_\mu,\crn
Y^-_\mu&=&-\sin\theta W'^-_\mu+\cos\theta Y'^-_\mu.\nn
\eea
In our model, the following limit is often taken into account:
\be \la^2_s,\la^2_\si, v^2_s,v^2_\si\ll u^2, u'^2, v^2, v'^2\ll
\om^2\sim \La^2_s\sim \La^2_\si. \label{limmit1}\ee

With the help of (\ref{limmit1}), the $\Ga$ in (\ref{Ga}) becomes
\be
\sqrt{\Ga} \simeq (4\La^2_s+2\La^2_\si+\om^2-u^2-u'^2)+\frac{32\La_s\La_\si v_s v_\si
+8\La_\si^2v^2_\si}
{4\La^2_s+2\La^2_\si+\om^2-u^2-u'^2},\label{Ga1}\ee
It is then
\be
M^2_W \simeq \frac{g^2}{2}\left(u^2+u'^2+v^2+v'^2+v^2_\rho\right)
-\frac{g^2}{2}\Delta_{M^2_\mathrm{w}},\label{MW}
\ee
with
\be
\Delta_{M^2_\mathrm{w}} = \frac{16\La_s\La_\si v_s v_\si
+4\La_\si^2v^2_\si}{4\La^2_s+2\La^2_\si+\om^2-u^2-u'^2}.\label{Deltaw}
\ee
Notice that in the limit $\la_{s}, \la_{\si}, v_{s}, v_\si \rightarrow 0$ then $\Ga
\simeq 4\La^2_s+2\La^2_\si+\om^2-u^2-u'^2$, the mixing angle $\theta$
tends to zero, and $M^2_W, M^2_Y$ in (\ref{MWY}) reduces to
\bea
M^2_W&=&\frac{g^2}{2}(u^2+u'^2+v^2+v'^2+v^2_\rho),\crn
M^2_Y&=&\frac{g^2}{2}\left(4\La^2_s+2\La^2_\si+\om^2+v^2
+v'^2+v^2_\rho\right), \label{MWYredu}
\eea
and one can evaluate
\be
\tan\theta \simeq  -\frac{8\La_sv_s+4\La_\si v_\si}{8\La^2_s+
4\La^2_\si+2\om^2}\sim \frac{v_s}{\La_s}\sim \frac{v_\si}{\La_\si},
\label{theta1}\ee
In addition, from (\ref{MWYredu}), it follows that  $M^2_W$ is much smaller than $M^2_Y$.

$\mathcal{L}^{NGB}_{mix}$ is the Lagrangian part of the neutral
gauge bosons $W_3, W_8, B, W_4$. The mass Lagrangian
in this case has the form
\bea
\mathcal{L}^{NGB}_{mix}
&=&\frac{(v^2+v'^2+v^2_\rho)}{324}\left(-9gW_{\mu3}+3\sqrt{3}gW_{\mu8}
+2\sqrt{6}g_X B_\mu\right)^2\crn
&+&\frac{\om^2}{108}\left(27g^2W^2_{\mu 4}+36g^2W^2_{\mu 8}
+12\sqrt{2}gg_XW_{\mu 8}B_\mu +2g^2_X B^2_\mu\right)\crn
&+&\frac{(u^2+u'^2)}{324}\left[81g^2W^2_{\mu 4}+(-9gW_{\mu3}-
3\sqrt{3}gW_{\mu8}+\sqrt{6}g_X B_\mu)^2\right]\crn
&+&\frac{g^2}{6}\left[2(2\La_s v_s+\La_\si v_\si)\left(3W_{\mu 3}W_{\mu 4}
-5\sqrt{3}W_{\mu 4}W_{\mu 8}
\right)+3(2v^2_s+v^2_\si+4\la^2_s+2\la^2_\si)W^2_{\mu3}\right.\crn
&+&\left.3(8v^2_s+4v^2_\si+2\la^2_s+\la^2_\si+2\La^2_s+\La^2_\si+
4\La_s\la_s+2\La_\si\la_\si)W^2_{\mu4}\right.\crn
&+&\left.2\sqrt{3}(-2v^2_s-v^2_\si+4\la^2_s+2\la^2_\si)W_{\mu3}W_{\mu8}
+(2v^2_s+v^2_\si+4\la^2_s+2\la^2_\si+16\La^2_s+
8\La^2_\si)W^2_{\mu8}\right.\crn
&+&\left.18(2\la_s v_s+\la_\si v_\si)W_{\mu3}W_{\mu4}+ 2\sqrt{3}(2
\la_s v_s+\la_\si v_\si)W_{\mu4}W_{\mu8}\right]\crn
&+&\frac{2t^2g^2}{27}(2\la^2_s+\la^2_\si+2\La^2_s
+\La^2_\si+4v^2_s+2v^2_\si)B^2_{\mu}
-\frac{2}{3}\sqrt{\frac{2}{3}}tg^2(2\la^2_s+\la^2_\si+2v^2_s
+v^2_\si)W_{\mu3}B_\mu\crn
&-&\frac{4}{3}\sqrt{\frac{2}{3}}tg^2\left[(2\la_s+2\La_s)v_s+(\la_\si
+\La_\si)v_\si\right]W_{\mu4}B_\mu \crn
&+&\frac{2\sqrt{2}}{9}tg^2(2v^2_s+v^2_\si+4\La^2_s+2\La^2_\si-
2\la^2_s-\la^2_\si)W_{\mu8}B_\mu.\label{LNGB}\eea
In the basis of $(W_{\mu3}, W_{\mu8}, B_{\mu}, W_{\mu4})$,
the $\mathcal{L}^{NGB}_{mix}$ in (\ref{LNGB}) can be rewritten:
\bea \mathcal{L}^{NGB}_{mix}&\equiv&\frac{1}{2}V^TM^2V, \label{MNGB1}\\
V^T&=&(W_{\mu3}, W_{\mu8}, B_{\mu}, W_{\mu4}),\crn
M^2&=&\frac{g^2}{4}\left(%
\begin{array}{cccc}
 M^2_{11}&M^2_{12}&M^2_{13}&M^2_{14} \\
 M^2_{12}&M^2_{22}&M^2_{23}&M^2_{24} \\
 M^2_{13}&M^2_{23}&M^2_{33}&M^2_{34} \\
 M^2_{14}&M^2_{24}&M^2_{34}&M^2_{44} \\
\end{array}%
\right),\label{Mmatrix}\eea
where
\bea
M^2_{11}&=&2\left(v^2+v'^2+u^2+u'^2+v^2_\rho+4v^2_s
+2v^2_\si+8\la^2_s+4\la^2_\si \right),\crn
M^2_{12}&=&-\frac{2\sqrt{3}}{3}\left(v^2+v'^2-u^2-u'^2+v^2_\rho
+4v^2_s+2v^2_\si-8\la^2_s-4\la^2_\si\right),\crn
M^2_{13}&=&-\frac{2}{3}\sqrt{\frac{2}{3}}t\left(2v^2+2v'^2+u^2+u'^2
+2v^2_\rho+8\la^2_s+4\la^2_\si+8v^2_s+4v^2_\si\right),\crn
M^2_{14}&=&8(3\la_s v_s+\La_s v_s)+4(3\la_\si v_\si+\La_\si v_\si),\crn
M^2_{22}&=&\frac{2}{3}\left(v^2+v'^2+4\om^2+u^2+u'^2
+v^2_\rho+4v^2_s+2v^2_\si+8\la^2_s+4\la^2_\si
+32\La^2_s+16\La^2_\si\right),\crn
M^2_{23}&=&\frac{2\sqrt{2}t}{9}\left(2v^2+2v'^2+2\om^2-u^2-u'^2
+2v^2_\rho+8v^2_s+4v^2_\si+16\La^2_s+8\La^2_\si-
8\la^2_s-4\la^2_\si\right),\crn
M^2_{24}&=&\frac{8}{\sqrt{3}}\left(\la_s v_s-5\La_s v_s\right)
+\frac{4}{\sqrt{3}}\left(\la_\si v_\si-5\La_\si v_\si\right),\crn
M^2_{33}&=&\frac{4t^2}{27}\left(4v^2+4v'^2+\om^2+u^2+u'^2
+4v^2_\rho+8\la^2_s+4\la^2_\si+8\La^2_s+4\La^2_\si
+16v^2_s+8v^2_\si \right),\crn
M^2_{34}&=&-\frac{32}{3}\sqrt{\frac{2}{3}}t\left(\la_s v_s
+\La_s v_s\right)-\frac{16}{3}\sqrt{\frac{2}{3}}t
\left(\la_\si v_\si+\La_\si v_\si\right) ,\crn
M^2_{44}&=&2\left(\om^2+u^2+u'^2+16v^2_s+8v^2_\si
+4\la^2_s+2\la^2_\si+4\La^2_s+2\La^2_\si
+8\La_s\la_s+4\La_\si\la_\si \right).
\label{elementsofM}\eea

The matrix $M^2$ in (\ref{Mmatrix})  with the elements
in (\ref{elementsofM}) has one exact eigenvalue, which is identified with
the photon mass,
\bea
M^2_{\ga}&=&0. \label{Mphoton}
\eea
The corresponding eigenvector of $M^2_{\ga}$ is
\bea
A_{\mu}&=&\left(%
\begin{array}{c}
  \frac{\sqrt{3}t}{\sqrt{4t^2+18}}\\
 -\frac{t}{\sqrt{4t^2+18}}\\
  \frac{3\sqrt{2}}{\sqrt{4t^2+18}}\\
  0\\
\end{array}%
\right).\label{Amu}\eea
Note that in the limit $\la_s, \la_\si, v_s, v_\si \rightarrow 0$,
$M^2_{14}=M^2_{24}=M^2_{34}=0$ and
$W_{4}$ does not mix with $W_{3\mu}, W_{8\mu}, B_\mu$. In the
general case $\la_s, \la_\si, v_s, v_\si \neq 0$,  the
 mass matrix in (\ref{Mmatrix})
contains  one exact eigenvalues as in (\ref{Mphoton}) with
the corresponding eigenstate being given in (\ref{Amu}).

The diagonalization of the mass matrix $M^2$ in (\ref{Mmatrix})
 is done via two steps. In
the first step, the basic $(W_{\mu3}, W_{\mu8}, B'_\mu, W_{4\mu})$
 is transformed into the basic
$(A_\mu, Z_\mu, Z'_\mu, W_{4\mu})$ by the matrix:
\be
U_{NGB} = \left(%
\begin{array}{cccc}
  s_W&-c_W&0&0\\
 -\frac{c_W t_W}{\sqrt{3}}&-\frac{s_W t_W}{\sqrt{3}}&\sqrt{1-\frac{t^2_W}{3}}&0\\
 c_W\sqrt{ 1 -\frac{t^2_W}{3}}&s_W\sqrt{ 1 -\frac{t^2_W}{3}}&\frac{t_W}{\sqrt{3}}&0\\
  0&0&0&1\\
\end{array}%
\right),\label{Us}
\ee
where $s_W=\sin\theta_W, c_W=\cos\theta_W, t_W=\tan\theta_W$, and we have
 used the continuation of the
gauge coupling constant $g$ of the $\mathrm{SU}(3)_L$ at the
spontaneous symmetry breaking point \cite{DongHLT, DongLongdh},
\be
t=\fr{3\sqrt{2}s_W}{\sqrt{3-4s^2_W}}.\label{t}\ee
The corresponding eigenstates are rewritten as follows
\bea A_\mu &=& s_W W_{3\mu}+c_W\left(-\fr{t_W}{\sqrt{3}}
W_{8\mu}+\sqrt{1-\fr{t^2_W}{3}}B_\mu\right),\crn
Z_\mu&=& - c_W W_{3\mu}+s_W\left(-\fr{t_W}{\sqrt{3}}
W_{8\mu}+\sqrt{1-\fr{t^2_W}{3}}B_\mu\right),\crn Z'_\mu &=&
\sqrt{1-\fr{t^2_W}{3}} W_{8\mu}+\fr{t_W}{\sqrt{3}}B_\mu.\eea
In this basis, the mass matrix $M^2$ becomes
\be M'^2=U^+_{NGB}M^2 U_{NGB}=\fr{g^2}{4}\left(%
\begin{array}{cccc}
  0 & 0 & 0 & 0 \\
  0 & M'^2_{22} & M'^2_{23} & M'^2_{24} \\
  0 & M'^2_{23} &M'^2_{33} &M'^2_{34} \\
  0 & M'^2_{24} & M'^2_{34} &M'^2_{44} \\
\end{array}%
\right).\label{M3}\ee
where
\bea
M'^2_{22} &=&\frac{4(2t^2+9)}{t^2+18}\left(8\la_s^2+4\la_\si^2
+u^2+u'^2+v^2+v'^2+v_\rho^2+4v_s^2+2v_\si^2\right),\crn
M'^2_{23} &=&\frac{4}{3\sqrt{3}}\frac{\sqrt{2t^2+9}}{{(t^2+18)}}
\left[(t^2-9)(8\la_s^2+4\la_\si^2+u^2+u'^2)+(2t^2+9)(v^2+v'^2
+v_\rho^2+4v_s^2+2v_\si^2)\right],\crn
M'^2_{24}&=&-4\sqrt{2}\sqrt{\frac{2t^2+9}{t^2+18}}\left[2(\La_s
+3\la_s)v_s+(\La_\si+3\la_\si) v_\si\right],\crn
M'^2_{33}&=&\frac{4}{27(t^2+18)}\left\{8\la^2_s(t^2-9)^2+8\La^2_s(t^2+18)^2
+81\left(4\la^2_\si+16\La^2_\si+4\om^2\right.\right.\crn
&+&\left.\left.u^2+u'^2+v^2+v'^2+v_\rho^2+4v_2^2+2v_\si^2\right)
+4\la^2_\si(t^2-18)t^2\right.\crn
&+&\left.t^2\left(144\La^2_\si+36\om^2-18u^2-18u'^2+36v^2
+36v'^2+36v^2_\rho+72v^2_s+36v^2_\si\right)
\right.\crn
&+&\left.t^4\left(4\La^2_\si+\om^2+u^2+u'^2+4v^2+4v'^2
+4v^2_\rho+16v^2_s+8v^2_\si\right)\right\},\crn
M'^2_{34}&=&-\frac{4\sqrt{2}}{3\sqrt{3}}\frac{1}{\sqrt{t^2+18}}
\left[(2\la_sv_s+\la_\si v_\si)(4t^2-9)+(2\La_sv_s+\La_\si v_\si)(4t^2+45)\right],\crn
M'^2_{44}&=&2\left(4\la^2_s+8\la_s\La_s+4\La_s^2
+2\la_\si^2+4\la_\si\La_\si+2\La_\si^2
+\om^2+u^2+u'^2+16v_s^2+8v_\si^2\right).\label{elementofM2}
\eea
In the approximation $\la^2_s\sim \la^2_\si, v^2_s, v^2_\si
\ll \La^2_s\sim \La^2_\si\sim \om^2$, we have
\bea
M'^2_{22} &=&\frac{2}{c_W^2}\left(u^2+u'^2+v^2+v'^2+v_\rho^2\right),\crn
M'^2_{23} &=&\frac{2[(1-2c^2_W)(u^2+u'^2)+v^2+v'^2
+v_\rho^2]\sqrt{\alpha_0}}{c^2_W },\crn
M'^2_{24}&=&-\frac{4}{c_W}\left(2\La_s v_s+\La_\si v_\si\right),\crn
M'^2_{33}&=&32(\La^2_\si+2\La^2_s) c^2_W\alpha_0
+8\om^2  c^2_W\alpha_0
+\frac{2}{ c^2_W}(v^2+v'^2+v^2_\rho)\alpha_0
+\frac{2}{ c^2_W}(2c^2_W-1)^2(u^2+u'^2)\alpha_0,\crn
M'^2_{34}&=&-\frac{8 x_0\sqrt{\al}}{c_W}\left(\La_s v_s
+4\La_\si v_\si\right),\crn
M'^2_{44}&=&2\left(\om^2+u^2+u'^2+4\La_s^2+2\La_\si^2
+8\la_s\La_s+4\la_\si\La_\si\right).\label{Mlimit}
\eea
with
\be
x_0 = 4c^2_W+1,\hs  \al_0=(4c^2_W-1)^{-1}.\label{x0al0}
\ee
It is noteworthy
that in the limit $v_s =0$ and $v_\si=0$, the elements
$M'^2_{24}$ and $M'^2_{34}$ vanish.
In this case the mixing between there is no mixing
 between $W_4$ and $Z_\mu, Z'_\mu$.

In the second step, three remain bosons gain masses via seesaw mechanism
\be
M^2_Z = \frac{g^2}{4}\left[M'^2_{22}-(M^{off})^T
 (M'^2_{2 \times 2})^{-1}M^{off}\right],\label{M2Z}
\ee
where
\be
M^{off} = \left(%
\begin{array}{c}
  M'^2_{23} \\
  M'^2_{24} \\
  \end{array}%
\right), \hs M'^2_{2\times2}=\left(%
\begin{array}{cc}
  M'^2_{33} &M'^2_{34} \\
  M'^2_{34} &M'^2_{44} \\
\end{array}%
\right).\label{Moff}
\ee
Combining (\ref{M2Z}) and (\ref{Moff}) yields:
\[
M^2_Z = \frac{g^2\left(u^2+u'^2+v^2+v'^2+v^2_\rho\right)}{2c^2_W}
-\frac{g^2}{2c^2_W}\Delta_{M^2_{z}},
\]
where
\be
\Delta_{M^2_{z}} = \frac{4\Delta^2_1}{x_2}+\frac{x_1(x_1 x_2-16
\Delta_1\Delta_2 x_0)}{x_2 (4c^4_W x_3+x_4)},\label{DeltaM}\ee
with
\bea
x_1&=&(1-2c^2_W)(u^2+u'^2)+v^2+v'^2+v^2_\rho,\crn
x_2&=&4\La_s(2\la_s+\La_s)+2\La_\si(2\la_\si+\La_\si)+\om^2+u^2+u'^2,\crn
x_3&=&8\La^2_s+4\La^2_\si+\om^2+u^2+u'^2,\crn
x_4&=&(1-4c^2)(u^2+u'^2)+v^2+v'^2+v^2_\rho,\crn
\Delta_1&=&2\La_s v_s+\La_\si v_\si,\hs
\Delta_2=\La_s v_s+4\La_\si v_\si.\label{x1234}
\eea
The $\rho$ parameter in our model is given by
\be
\rho = \frac{M^2_W}{M^2_Z \cos^2\theta_W}
=1+\frac{\delta_{\mathrm{w}z}}
{M^2_{z}}\equiv 1+\delta_{\mathrm{tree}}.\label{rhoparametre}
\ee
where
\be
\delta_{\mathrm{w}z} = \frac{g^2}{2c^2_W}
\left(\Delta_{M^2_{z}}-\Delta_{M^2_\mathrm{w}}\right).\label{Deltazw}
\ee
From the mass of $W$ boson evaluated in (\ref{MWYredu}) we can identify
\[
2(u^2+u'^2+v^2+v'^2+2v^2_\rho)=v^2_{weak}\simeq (246\,  \mathrm{GeV})^2\]
and then obtain \be
u\sim u' \sim v'\sim v\sim 100\,  \mathrm{GeV},\label{uupvvp}\ee
provided that $v_\rho \sim 0$.

\begin{figure}[h]
\begin{center}
\includegraphics[width=12.0cm, height=4.5cm]{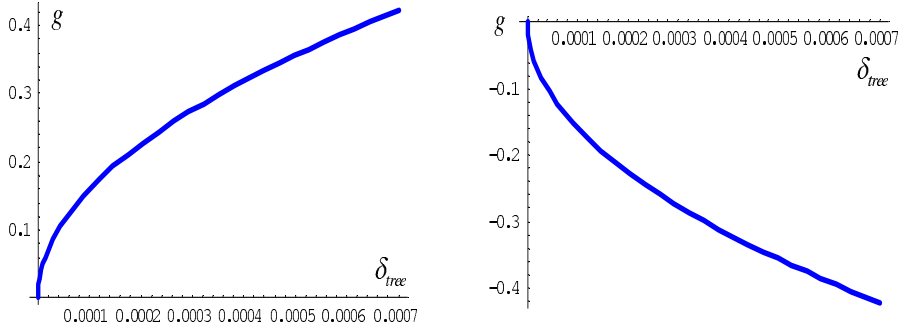}
\vspace*{-0.4cm}

 \caption[The coupling $g$ as a function of
$\delta_{\mathrm{tree}}$ with $\delta_{\mathrm{tree}} \in (0,
0.0007)$ and $v_{s} = 10 \, \mathrm{GeV}$]{The coupling  $g$ as a
function of  $\delta_{ \mathrm{tree}}$ with
$\delta_{\mathrm{tree}}\in (0, 0.0007)$ and $v_{s} = 10\,
\mathrm{GeV}$}\label{gdrelat}
\end{center}
\end{figure}

In addition, let us assume the relations (\ref{limmit1}) and put
\bea
\la_s&=&v^2_s/\La_s= v^2_\si/\La_\si =\la_\si, \crn
\om&=&\La_s\equiv \La_\si \nn \eea
then
\be
\Delta_{M^2_{z}}-\Delta_{M^2_\mathrm{w}}\simeq
\frac{(2.28571\La^2_s-45590.8)v^2_s+2.77644\times 10^{6}}{\La^2_s}.\label{deltae}
\ee
From (\ref{rhoparametre}) -- (\ref{deltae}) we have:
\be
\delta_{\mathrm{tree}} = \frac{g^2}{2c^2_W}\frac{(2.28571
\La^2_s-45590.8)v^2_s+2.77644\times 10^{6}}{M^2_{z} \La^2_s}.\label{Delta1}
\ee
The experimental value of the $\rho$ parameter
and $M_W$ are respectively given in Ref. \cite{PDG2012}
\bea
\rho&=&1.0004^{+0.0003}_{-0.0004}\hs\hs (\delta_{\mathrm{tree}}=0.0004^{+0.0003}_{-0.0004}),\crn
s^2_W&=&0.23116 \pm 0.00012,\crn
M_W&=&80.358\pm 0.015 \, \mathrm{GeV}.\label{rhoMwMz12}
\eea
It means
\bea
0  \leq  \delta_{\mathrm{tree}} \leq 0.0007.\label{rhotree}
\eea
The expression (\ref{Delta1}) gives the relations between $g$ and $\mathrm{\delta_{tree}}$ as follows
\[
g = \pm \frac{79.9648 \sqrt{\mathrm{\delta_{tree}}}v^2_s}{
\sqrt{1.14286 v^6_s-22795.4v^2_s-45590.8\times 10^{6}}}.
\]
In the Fig. \ref{gdrelat} we have plotted  $g$ as a function of  $\delta_{\mathrm{tree}}\in (0, 0.0007) $
from which it  provides that $v_s=10 \mathrm{GeV}$  satisfying  the condition (\ref{limmit1}).
From the Fig. \ref{gdrelat}, we can find out $|g| \in (0, 0.42)$.

Diagonalizing the mass matrix $M'^2_{2\times 2}$, we get two new physical gauge bosons
\bea
Z''_\mu&=&\cos\phi Z'_\mu+\sin\phi W_{\mu4},\crn
W'_{\mu4}&=&-\sin\phi Z'_\mu+\cos\phi W_{\mu4}.\label{ZppW4p}
\eea
The mixing angle $\phi$ is given by
\be
\tan\phi = \frac{8\sqrt{\al_0}c_W(\La_s v_s+4\La_\si v_\si)x_0}{-
4c^4_W\al_0x_3+c^2_W x_2-\al_0 x_4+\sqrt{F}}.
\label{tanphi}\ee
where
\bea
F&=&\left(4c^4_W\al_0x_3+c^2_Wx_2+\al_0 x_4\right)^2-4\al_0c^2_W\left[(u^2+u'^2)x_5
-4c^2_W(u^2+u'^2)x_2+4c^4_Wx_3 x_2\right.\crn
&+&\left. (v^2+v'^2+v^2_\rho)x_5+8\la_s\La_s(u^2+u'^2+v^2+v'^2
+v^2_\rho)-16(\La_s v_s+4\La_\si v_\si)^2x^2_0\right], \label{F}
\eea
and
\[
x_5 = 4\La^2_s+4\La_\si\la_\si+2\La^2_\si+\om^2+u^2+u'^2,
\]
 With the help of (\ref{limmit1}),  we have
 \[
 F\simeq c^2_W\left\{8\La_s\la_s+2\La_\si(2\la_\si+\La_\si)+
 \om^2-4\left[(8\al_0 c^2_W-1)\La^2_s+\al_0 c^2_W(4\La^2_\si+\om^2)\right]\right\}\]
  and one can evaluate
\be \tan\phi \simeq \frac{4\sqrt{\al_0}(\La_sv_s+4\La_\si
v_\si)x_0}{c_W\left[(4-32\al_0 c^2_W)\La^2_s+ (2-16\al_0
c^2_W)\La^2_\si+(1-4\al_0 c^2_W)\om^2\right]} \sim
\frac{v_s}{\La_s}\sim \frac{v_\si}{\La_\si}. \label{tanphi1}\ee
The physical mass eigenvalues are defined by \be
M^2_{Z''_\mu,W'_{\mu4}}=\frac{g^2}{4c^2_W}\left\{4\al_0c^4_Wx_3
+c^2_Wx_2+ax_4\pm\sqrt{F}\right\}. \label{ZppW4p}\ee In the limit
$\la_{s}, \la_{\si}, v_{s}, v_\si \rightarrow 0$ the mixing angle
$\phi$ tends to zero, and $M^2_{Z''_\mu,W'_{\mu4}}$ in
(\ref{ZppW4p}) reduces to \bea
M^2_{Z''_\mu}&=&\frac{g^2}{2c^2_W}\left(x_4+4c^2_W x_3\right),\crn
M^2_{W'_{\mu4}}&=&\frac{g^2}{2}\left(u^2+u'^2+\om^2+4\La^2_s
+2\La^2_\si\right),\label{MZW4redu} \eea Thus the $W'_{\mu4}$ and
$W_5$ components have the same mass. With this result, we should
identify the combination of $W'_{\mu4}$ and $W_5$ \be
\sqrt{2}X^0_\mu = W'_{\mu4}-iW_5 \label{Xmu} \ee as physical
neutral non-Hermitian gauge boson. The superscript ``0''
denotes neutrality of gauge boson $X$. Notice that, the
identification in (\ref{Xmu}) only can be acceptable with the
limit $\la_{s,\si}, v_{s,\si} \rightarrow 0$. In general it is not
true because of the difference in masses of $W'_{\mu4}$ and
$W_{\mu5}$ as in (\ref{mW5}) and (\ref{ZppW4p}).

The expressions (\ref{theta1}) and (\ref{tanphi1}) show that, with the limit (\ref{limmit1}),
the mixings between the charged gauge bosons $W-Y$ and
 the neutral ones $Z'-W_4$ are in the same
order since they are proportional to $v_s/\La_s$ (or $v_\si/\La_\si$).
In addition, from (\ref{MWYredu})
$M^2_{Z''_\mu}\simeq 2g^2(8\La^2_s+4\La^2_\si+\om^2)$ is little
bigger than $M^2_{W'_{\mu4}}\simeq
\frac{g^2}{2}\left(\om^2+4\La^2_s+2\La^2_\si\right)$ \,\,(or $M^2_{X^0_{\mu}}$),
 and $|M^2_Y-M^2_{X^0_{\mu}}|=
\frac{g^2}{2}(v^2+v'^2+v^2_\rho+u^2+u'^2)$ is litle smaller than
$M^2_W=\frac{g^2}{2}(u^2+u'^2+v^2+v'^2+v^2_\rho)$.
In that limit, the masses of $X^0_{\mu}$ and $Y$ is nearly degenerate.

\section{\label{conclus}Conclusions}

In this paper, we have constructed the $D_4$ model based on
$\mathrm{SU}(3)_C \otimes \mathrm{SU}(3)_L \otimes
\mathrm{U}(1)_X$ gauge symmetry responsible for fermion masses and
mixing. Neutrinos get masses from antisextets which is in a
singlet and a doublet under $D_4$. We argue how flavor mixing
patterns and mass splitting are obtained with a perturbed $D_4$
symmetry. We have pointed out that this model is more simpler than
those of $S_3$ and $S_4$ \cite{dlnvS3, dlsvS4} since the same
number of Higgs multiplets are needed in order to allow the
fermions to gain masses
 but with the simple scalar Higgs potential.
The CKM  matrix is the identity matrix at the tree-level, but it
can be different from it by adding the soft violating terms. The realistic neutrino mixing, by old
data with $\theta_{13}=0$, can be obtained only if  the direction
for breaking $D_4 \rightarrow Z_2$. For the case with the nonvanishing $\theta_{13}$, it is
necessary to introduce one more Higgs triplet $\rho$ which is in
$1^{\prime \prime \prime}$ of the $D_4$ group responsible
for breaking the $Z_2\rightarrow
\{\mathrm{identity}\}$. As a result, the
value of $\theta_{13}$ is a small perturbation by
$\frac{v_\rho}{\La}$.

\section*{Acknowledgments}
This research is funded by Vietnam National Foundation for Science
and Technology Development (NAFOSTED) under grant number
103.01-2011.63.

\appendix
\section{\label{apa}$\emph{D}_4$ group and Clebsch-Gordan coefficients}
$D_4$ is the symmetry group of a square \cite{nhat}. It has eight
elements divided into five conjugacy classes, with $\underline{1},
\underline{1}', \underline{1}'', \underline{1}'''$ and
$\underline{2}$ as its five irreducible representations. Any
element of $D_4$ can be formed by multiplication of the generators
$a$ (the $\pi/2$ rotation) and $b$ (the reflection) obeying the
relations $\emph{a}^4=\emph{e}$, $\emph{b}^2=\emph{e}$, and
$\emph{bab}=\emph{a}^{-1}$. $D_4$ has the following five conjugacy
classes,
\begin{eqnarray}
C_1 &:& \{a_1\equiv e\},
\crn
C_2 &:& \{a_2\equiv a^2\},
\crn
C_3 &:& \{a_3\equiv a,\,a_4\equiv a^3\},
\\
C_4 &:& \{a_5\equiv b,\,a_6\equiv a^2b\},
\crn
C_5 &:& \{a_7\equiv ab,\,a_8\equiv a^3b\}.\nn
\end{eqnarray}

The character table of $D_4$ is given as follows:
\bc
\begin{tabular}{|c||c|c|c|c|c|c|c|}
\hline \,Class & $\,n$ & $\,h$ & $\,\chi_1$ & $\,\chi_{1'}$ &
$\,\chi_{1''}$& $\,\chi_{1'''}$& $\,\chi_2$
\\\hline \hline
$C_1$ & 1 & 1 & 1 & 1 & 1 & 1 &2\\
$C_2$ & 1 & 2 & 1 & 1 & 1 &1 &-2\\
$C_3$ & 2 & 4 & 1 & -1& -1& 1 &0\\
$C_4$ & 2 & 2 & 1 & 1 & -1& -1 &0\\
$C_5$ & 2 & 2 & 1 & -1& 1 &-1 &0\\
\hline
\end{tabular}
\ec
where $n$ is the order of class and $h$ the order of elements
within each class.

We have worked in real basis, in which the two - dimensional
representation $\underline{2}$ of $D_4$ is real, $2^*(1^*,
2^*)=2(1^*, 2^*)$. One possible choice of generators is given as
follows \bea \underline{1}&:& a=1,\hs b=1, \crn \underline{1}' &:&
a=1,\hs b=-1,\crn \underline{1}'' &:& a=-1,\hs b=1, \crn
\underline{1}''' &:& a=-1,\hs b=-1,\crn
\underline{2} &:& a=\left(%
\begin{array}{cc}
  0 & 1 \\
  -1 & 0 \\
\end{array}%
\right),\hs b=\left(%
\begin{array}{cc}
  1 & 0 \\
  0 & -1 \\
\end{array}%
\right).\eea Using them we calculate the Clebsch-Gordan
coefficients for all the tensor products as given below.

First, let us put $\underline{2}(1,2)$ which means some
$\underline{2}$ doublet such as $x=(x_1,x_2)\sim \underline{2}$ or
$y=(y_1,y_2)\sim \underline{2}$ and so on, and similarly for the
other representations. Moreover, the numbered multiplets such as
$(...,ij,...)$ mean $(...,x_i y_j,...)$ where $x_i$ and $y_j$ are
the multiplet components of different representations $x$ and $y$,
respectively. In the following the components of representations
in left-hand side will be omitted and should be understood, but
they always exist in order in the components of decompositions in
right-hand side:
 \bea
\underline{1}(1)\otimes\underline{1}(1)&=&\underline{1}(11),\hs\underline{1}'(1)\otimes
\underline{1}'(1)=\underline{1}(11),\crn
\hs\underline{1}{''}(1)\otimes \underline{1}{''}(1)&=&\underline{1}(11),
\hs\underline{1}{'''}(1)\otimes \underline{1}{'''}(1)=\underline{1}(11),\\
\underline{1}(1)\otimes\underline{1}'(1)&=&\underline{1}'(11),
\hs\underline{1}(1)\otimes\underline{1}{''}(1)=\underline{1}{''}(11),\crn
\underline{1}(1)\otimes\underline{1}{'''}(1)&=&\underline{1}{'''}(11),\hs
\underline{1}'(1)\otimes\underline{1}{''}(1)=\underline{1}{'''}(11),\crn
\underline{1}{''}(1)\otimes\underline{1}{'''}(1)&=&\underline{1}{'}(11),\hs
\underline{1}{'''}(1)\otimes\underline{1}{'}(1)=\underline{1}{''}(11),\\
\underline{1}(1)\otimes\underline{2}(1,2)&=&\underline{2}(11,12),\hs
\underline{1'}(1)\otimes\underline{2}(1,2)=\underline{2}(11,-12),\crn
\underline{1}{''}(1)\otimes\underline{2}(1,2)
&=&\underline{2}(12,11),\hs
 \underline{1}{'''}(1)\otimes\underline{2}(1,2)=\underline{2}(-12,11),\\
\underline{2}(1,2) \otimes \underline{2}(1,2)
&=&\underline{1}(11+22) \oplus \underline{1}'(11-22) \oplus
\underline{1}{''}(12+21)\oplus \underline{1}{'''}(12-21).\eea In
the text we usually use the following notations, for example,
$(xy)_{\underline{1}}\equiv(x_1y_1+x_2y_2)$ which is the
Clebsch-Gordan coefficients of $\underline{1}$ in the
decomposition of $\underline{2}\otimes \underline{2}$, where as
mentioned $x=(x_1,x_2)\sim \underline{2}$ and $y=(y_1, y_2)\sim
\underline{2}$.

The rules to conjugate the representations \underline{1},
\underline{1}$'$, \underline{1}$''$, \underline{1}$'''$ and
\underline{2} are given by \bea
\underline{2}^*(1^*,2^*)&=&\underline{2}(1^*, 2^*),\\
\underline{1}^*(1^*)&=&\underline{1}(1^*),\,\,\,
\underline{1}'^*(1^*)=\underline{1}'(1^*),\,\,\,
\underline{1}''^*(1^*)=\underline{1}''(1^*),\,\,\,
\underline{1}'''^*(1^*)=\underline{1}'''(1^*),\eea where, for
example, $\underline{2}^*(1^*,2^*)$ denotes some $\underline{2}^*$
multiplet of the form $(x^*_1,x^*_2)\sim
\underline{2}^*$.

\section{\label{apb}The numbers}
In the following we will explicitly point out the lepton number
($L$) and lepton parity ($P_l$) of the model particles (notice
that the family indices are suppressed): \bc
\begin{tabular}{|c|c|c|}
  \hline
Particles & $L$ & $P_l$  \\ \hline
 $N_R$, $u$, $d$,  $\phi^+_1$,$\phi'^+_1$, $\phi^0_2$,$\phi'^0_2$,
  $\eta^0_1$,$\eta'^0_1$, $\eta^-_2$,$\eta'^-_2$
  $\chi^0_3$, $\sigma^0_{33}$, $s^0_{33}$ & 0 & 1  \\ \hline
  $\nu_L$, $l$, $U$, $D^*$, $\phi^+_3$,$\phi'^+_3$, $\eta^0_3$,
  $\eta'^0_3$, $\chi^{0*}_1$, $\chi^+_2$,
   $\sigma^0_{13}$,
   $\sigma^+_{23}$, $s^0_{13}$, $s^+_{23}$ & $-1$ & $-1$
   \\ \hline
   $\sigma^{0}_{11}$, $\sigma^{+}_{12}$, $\sigma^{++}_{22}$,
   $s^{0}_{11}$, $s^{+}_{12}$, $s^{++}_{22}$ & $-2$ & 1 \\ \hline
\end{tabular}\ec

\end{document}